\documentclass[a4paper,fleqn]{cas-dc}

\usepackage[numbers]{natbib}
\usepackage{amssymb}
\usepackage{xcolor}
\usepackage{multirow}
\usepackage{colortbl}
\usepackage{booktabs}
\usepackage{arydshln}
\usepackage{caption}
\usepackage{multicol}
\usepackage{subcaption}

\definecolor{myPurple}{RGB}{128,0,128}
\definecolor{myBlue}{RGB}{0,176,240}
\definecolor{myGreen}{RGB}{0,176,80}
\definecolor{myRed}{RGB}{255,0,0}
\definecolor{darkgreen}{rgb}{0,0.5,0}
\definecolor{myGray}{RGB}{175,171,171}
\definecolor{myOrange}{RGB}{255,192,0}

\newcommand{\apicoder}{APICoder\xspace}
\newcommand{\apifinder}{APIFinder\xspace}
\newcommand{\pandaseval}{PandasEval\xspace}
\newcommand{\numpyeval}{NumpyEval\xspace}
\newcommand{\monkeyeval}{MonkeyEval\xspace}
\newcommand{\beatnumeval}{BeatNumEval\xspace}
\newcommand{\torchdataeval}{TorchDataEval\xspace}
\newcommand{\torchdatacomplexeval}{TorchDataComplexEval\xspace}
\newcommand{\codegen}{\textsc{CodeGen}\xspace}

\newcommand{\codegenapi}{\textsc{CodeGenAPI}\xspace}
\newcommand{\apidoc}{API documentation\xspace}
\newcommand{\passk}[0]{${\rm{pass}}@{k}$ }
\newcommand{\Passk}[0]{${\rm{Pass}}@{k}$ }

\def\tsc#1{\csdef{#1}{\textsc{\lowercase{#1}}\xspace}}
\tsc{WGM}
\tsc{QE}
\tsc{EP}
\tsc{PMS}
\tsc{BEC}
\tsc{DE}

\begin{document}

\let\WriteBookmarks\relax
\def\floatpagepagefraction{1}
\def\textpagefraction{.001}
\shorttitle{Private-Library-Oriented Code Generation with Large Language Models}
\shortauthors{Daoguang Zan et~al.}

\title [mode = title]{Private-Library-Oriented Code Generation with Large Language Models}




\author[1]{Daoguang Zan}[]
\fnmark[1]
\ead{daoguang@iscas.ac.cn}
\credit{Conceptualization, Data curation, Formal analysis, Investigation, Methodology, Resources, Writing - original draft}
\address[1]{Institute of Software, Chinese Academy of Sciences, Beijing, China}

\author[2]{Bei Chen}[]
\cormark[1]
\ead{beichen@microsoft.com}
\credit{Conceptualization, Formal analysis, Methodology, Supervision, Writing - original draft}
\address[2]{Microsoft, Beijing, China}

\author[3]{Yongshun Gong}[]
\cormark[1]
\ead{ysgong@sdu.edu.cn}
\credit{Conceptualization, Methodology, Validation}
\address[3]{School of Software, Shandong University, Jinan, China}

\author[4]{Junzhi Cao}[]
\credit{Conceptualization, Methodology, Validation}
\address[4]{New York University, New York, USA}

\author[5]{Fengji Zhang}[]
\credit{Data curation, Formal analysis, Investigation}
\address[5]{School of Computer Science, Wuhan University, Wuhan, China}

\author[1]{Bingchao Wu}[]
\credit{Resources, Visualization}

\author[1]{Bei Guan}[]
\cormark[1]
\ead{guanbei@iscas.ac.cn}
\credit{Methodology, Validation}

\author[3]{Yilong Yin}[]
\credit{Conceptualization, Methodology, Supervision, Project administration}

\author[1]{Yongji Wang}[]
\credit{Conceptualization, Methodology, Supervision, Project administration}






\cortext[cor1]{Corresponding author}


\begin{abstract}
Large language models (LLMs), such as Codex and GPT-4, have recently showcased their remarkable code generation abilities, facilitating a significant boost in coding efficiency.
This paper will delve into utilizing LLMs for code generation in private libraries, as they are widely employed in everyday programming.
Despite their remarkable capabilities, generating such private APIs poses a formidable conundrum for LLMs, as they inherently lack exposure to these private libraries during pre-training.
To address this challenge, we propose a novel framework that emulates the process of programmers writing private code.
This framework comprises two modules:
\apifinder first retrieves potentially useful APIs from \apidoc; and \apicoder then leverages these retrieved APIs to generate private code.
Specifically, \apifinder employs vector retrieval techniques and allows user involvement in the retrieval process.
For \apicoder, it can directly utilize off-the-shelf code generation models.
To further cultivate explicit proficiency in invoking APIs from prompts, we continuously pre-train a reinforced version of \apicoder, named \codegenapi.
Our goal is to train the above two modules on vast public libraries, enabling generalization to private ones.
Meanwhile, we create four private library benchmarks, including \torchdataeval, \torchdatacomplexeval, \monkeyeval, and \beatnumeval, and meticulously handcraft test cases for each benchmark to support comprehensive evaluations.
Numerous experiments on the four benchmarks consistently affirm the effectiveness of our approach. Furthermore, deeper analysis is also conducted to glean additional insights.
\end{abstract}


\begin{highlights}
\item We introduce a new scenario focusing on private-library-oriented code generation and propose an innovative framework miming the human process of using private libraries. This framework consists of two main modules: \apifinder and \apicoder.
\item To fairly evaluate this scenario, we manually construct four benchmarks, namely \torchdataeval, \torchdatacomplexeval, \monkeyeval, and \beatnumeval, each featuring an extensive array of test cases.
\item Our comprehensive experiments exhibit the superior performance of our proposed framework, showcasing its effectiveness and efficiency in handling private APIs.
\end{highlights}


\maketitle

\section{Introduction}

\begin{figure}
	\centering
	\includegraphics[scale=.48]{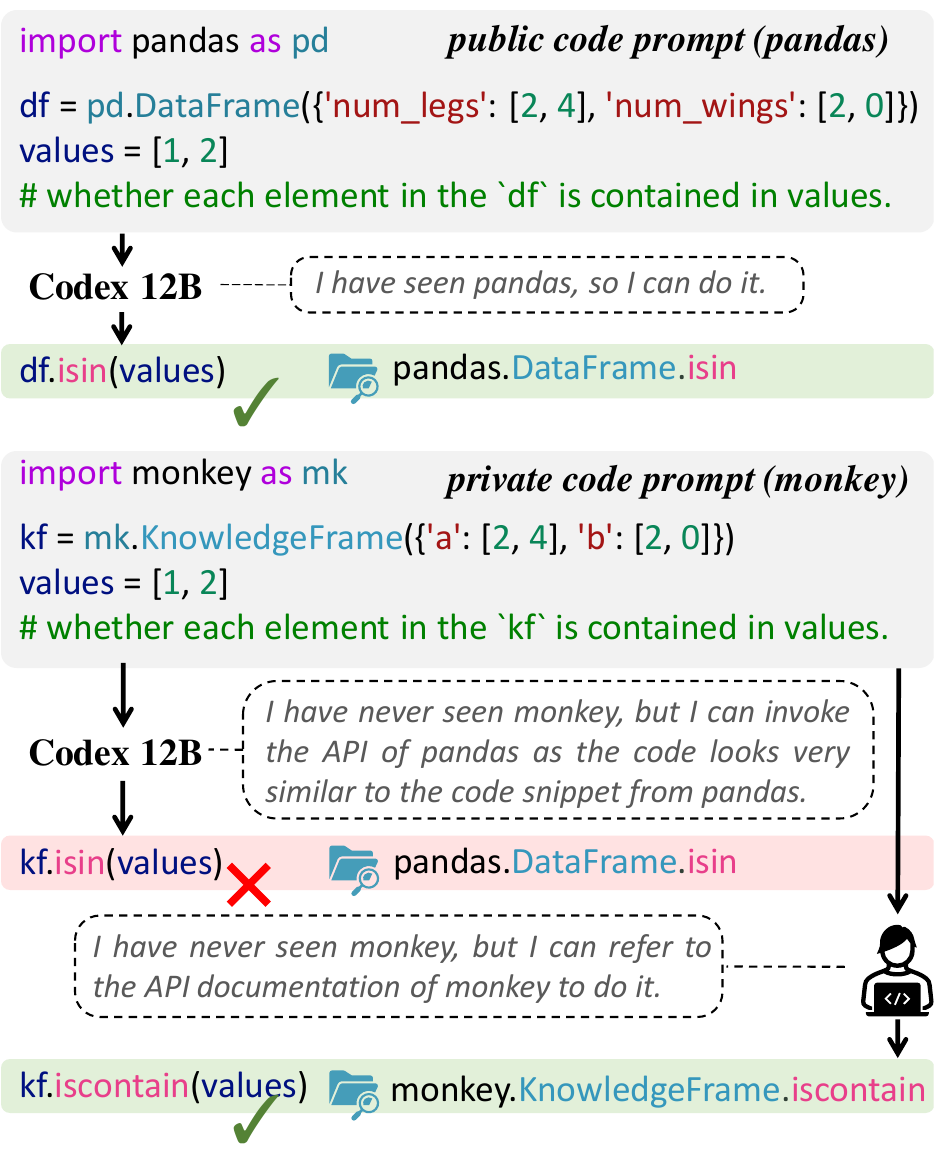}
        \caption{An example of code generation on public library (pandas) and private library (monkey) by Codex $12$B and a programmer.}
	\label{fig:example_llm_private}
\end{figure}

The use of third-party code libraries can greatly enhance code reusability, reduce development difficulty, and accelerate development speed in software development~\cite{third-part-lib-3,third-part-lib-2,third-part-lib-1}.
Developers can expedite their development process by integrating third-party libraries, while avoiding the redundant work of reinventing the wheel.
Considering the many advantages of using third-party code libraries, it is common for companies to develop and maintain a code library aimed at better code control and version management.
However, in contrast to public code libraries, many code libraries maintained by companies are private and exclusively used for internal purposes, with the goal of enhancing security measures for the protection of intellectual property and trade secrets~\cite{private-paper-2}.
Hence, we can see that leveraging private libraries for code generation is a prevalent and meaningful scenario~\cite{private-paper-1}.
With the swift advancement of large language model (LLM), this technology has been applied in code generation field~\cite{codet,nl2code-survey-2,nl2code-survey} via training on extensive code corpora, such as Codex~\cite{codex}, AlphaCode~\cite{alphacode}, \codegen~\cite{codegen}.
Unfortunately, despite these language models possessing incredibly impressive code generation abilities in general scenarios, they are powerless when handling private libraries as they have no prior knowledge of such information.
Figure~\ref{fig:example_llm_private} gives a concrete example to illustrate it.
When generating code for the public library pandas, Codex, a $12$B large language model, successfully invokes the library's API calls and produces effective code.
When we encapsulate the \texttt{pandas.DataFrame.isin} into \texttt{monkey.KnowledgeFrame.iscontain} for internal usage; however, Codex is unable to correctly invoke the API and generate code for the private library by referencing \apidoc, as a programmer would.
This example further underscores the challenges of LLMs in facing private-library-oriented code generation.

In this paper, our proposal seeks to tackle these challenges by introducing a framework that unleashes large language models with the aptitude to produce code that invokes private libraries.
As private libraries often feature comprehensive reference materials such as \apidoc, our core idea involves emulating the process by which programmers use \apidoc to develop code for private libraries.
The process of searching information in \apidoc to understand how to use an API before writing code, commonly referred to as \apidoc lookup~\cite{apilookup}, is an essential practice in software development and programming.
Likewise, our framework consists of two modules: \apifinder retrieves potential APIs from the \apidoc based on user requirements, and \apicoder leverages the retrieved APIs to produce code.
Regarding \apifinder, we employ a dense retriever and incorporate a user-friendly interface that provides optional user engagement in the API retrieval process.
Regarding \apicoder, off-the-shelf language models for code generation like \codegen can be utilized to write private code directly.
Even more thrillingly, we discover that continuously pre-training the existing models using our proposed novel strategies can significantly improve their capability of calling private library APIs.
Since we are unable to access the corpus of private libraries, we train both \apifinder and \apicoder on many crawled public libraries with the aspiration of enabling generalization to private library scenarios.

To the best of our knowledge, we are the first to propose the private-library-oriented code generation scenario.
Due to the lack of available benchmarks for evaluating this scenario, we create four benchmarks: \torchdataeval, \torchdatacomplexeval, \monkeyeval, and \beatnumeval.
Both \torchdataeval and \torchdatacomplexeval consist of $50$ programming problems for private libraries, with the latter featuring more challenging and practical problems.
\monkeyeval and \beatnumeval, each containing $101$ programming problems, are adaptations of PandasEval and NumpyEval~\cite{CERT}, respectively.

This paper serves as an extended edition of~\cite{apicoder} by enhancing various aspects of the original work. Specifically, these aspects can be summarized as follows:
(1) We investigate private-library-oriented code generation across varying levels of difficulty and further craft a more challenging and real-world private library benchmark named \torchdatacomplexeval.
(2) We explore the impact of all components of \apidoc, such as API examples and parameters, on the effectiveness of LLMs.
(3) We conduct extensive experiments on $17$ popular code generation models, and further train \codegenapi using three prompts across various model sizes ($350$M, $2$B, $6$B).
(4) We perform a more comprehensive and rigorous analysis and evaluation, yielding numerous intriguing and valuable insights.
Overall, our contributions are listed as follows:
\begin{itemize}
    \item We introduce a scenario called private-library-oriented code generation and propose a simple yet innovative framework that emulates the human approach to utilizing private libraries.
    \item We meticulously develop four private library benchmarks, each equipped with a comprehensive suite of test cases.
    \item Thorough experiments highlight the remarkable performance of our framework, emphasizing its feasibility in invoking private APIs.
\end{itemize}

\section{Related Work}
\subsection{Large Language Models for Code Generation}
Pioneering models like PLBART~\cite{plbart}, PyMT5~\cite{pymt5}, and GPT-C~\cite{gpt-c} possess modest parameter counts and exhibit limited proficiency in zero-shot code generation. 
In contrast, although large-scale models such as GPT-Neo~\cite{gpt-neo} and GPT-J~\cite{gpt-j} boast billions of parameters, their effectiveness in the code generation task remains constrained by the scarcity of code within their training datasets.
Lately, a variety of advanced LLMs have been introduced for code generation, including but not limited to Codex~\cite{codex}, PaLM-Coder~\cite{palm}, PanGu-Coder~\cite{pangu-coder,pangucoder2}, AlphaCode~\cite{alphacode}, GPT-4~\cite{gpt4}, and ChatGPT\footnote{\url{https://chat.openai.com}}. 
These models encompass extensive parameter counts and are trained on premium code-rich data. Although they exhibit outstanding performance in code generation tasks, the majority of them are inaccessible.
Currently, numerous outstanding publicly available models have emerged, such as CodeParrot~\cite{codeparrot}, CodeGen~\cite{codegen}, and PolyCoder~\cite{polycoder}. These models play a vital role in the progress of LLMs for code generation.
While all of the above models only support code generation from left to right, models such as SantaCoder~\cite{santacoder}, FIM~\cite{fim}, InCoder~\cite{incoder}, StarCoder~\cite{starcoder}, and MIM~\cite{mim} also support inserting code snippet at arbitrary positions, including intermediate locations.
Besides Python files, JuPyT5~\cite{jupyt5} opts for Jupyter Notebook files for training and achieves noteworthy results.
Recent models, such as ERNIE-Code~\cite{ernie-code}, BLOOM~\cite{bloom}, and CodeGeeX~\cite{codegeex}, have also considered a new scenario of multiple programming or natural languages.
In this study, we will delve into how to equip the aforementioned models with the capability to utilize private libraries.

In the era of LLMs, hand-crafted code generation benchmarks play a pivotal role in ensuring that the programming problems have not been exposed during pre-training.
Consequently, HumanEval~\cite{codex} and MBPP~\cite{mbpp} were proposed and gained widespread popularity.
The above two benchmarks are English and Python versions.
Subsequently, several benchmarks extended them to multiple languages, such as MBXP~\cite{mbxp}, MultiPL~\cite{multipl-e}, and HumanEval-X~\cite{codegeex}.
Furthermore, recently released benchmarks are increasingly tailored to some real-world scenarios.
For example, APPs~\cite{apps} and CodeContests~\cite{alphacode} served as benchmarks for assessing programming contest abilities;
DSP~\cite{dsp} and DS-1000~\cite{ds-1000} are benchmarks oriented towards data science;
MTPB~\cite{codegen} was introduced for multi-turn dialogue code generation.
SecurityEval~\cite{securityeval} was proposed for evaluating code security;
MathQA-Python~\cite{mbpp} and GSM8K-Python~\cite{GSM8K-Python} are geared towards evaluating mathematical proficiency;
PandasEval~\cite{jigsaw,CERT} and NumpyEval~\cite{CERT} test code generation in public libraries, while some benchmarks~\cite{cocomic,repository,repocoder} concentrate on the repository or cross-file level.
In this paper, we propose a private library scenario for code generation and meticulously craft four benchmarks, named \torchdataeval, \torchdatacomplexeval, \monkeyeval, and \beatnumeval.

\subsection{Retrieval-based Generation}
Retrieval-based generation~\cite{formal2022distillation,santhanam2021colbertv2,xiong2020approximate} has garnered significant popularity in the fields of natural language processing and information retrieval, owing to its adeptness at capitalizing on externally pre-existing knowledge.
For instance, Fusion-in-Decoder~\cite{fusion_in_decoder}, DPR~\cite{dense-retrieval}, and RocketQA~\cite{rocketqa} leverage this technique for open-domain question answering by retrieving relevant passages and exhibit remarkable results.
In the field of code generation, there also exist some efforts utilizing retrieval techniques, such as DeepAPI~\cite{deep_api_learning}, ReACC~\cite{retri-1}, REDCODER~\cite{redcoder}, RepoCoder~\cite{repocoder}, and DocCoder~\cite{retri-2}.
In this paper, our goal is to retrieve real-world private APIs from \apidoc.

\section{Preliminaries}
\subsection{Task Definition}
\begin{figure}
	\centering
	\includegraphics[scale=0.5]{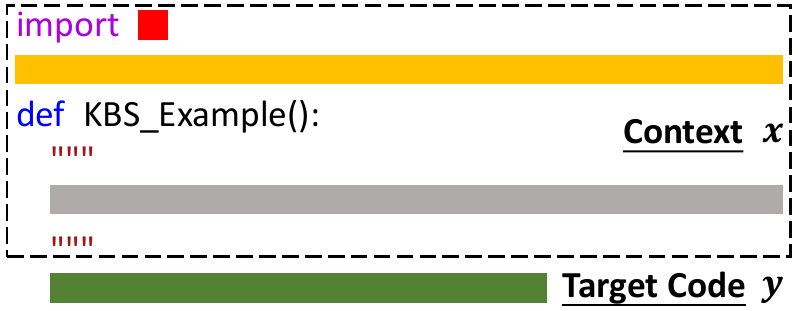}
        \caption{One simple Python example of code generation. \textcolor{myRed}{$\blacksquare$} represents code library likes pandas; \textcolor{myOrange}{$\blacksquare$} represents some code snippets, functions, or classes; \textcolor{myGray}{$\blacksquare$} represents a natural language description of programming problem; \textcolor{darkgreen}{$\blacksquare$} represents target code that solves the programming problem in \textcolor{myGray}{$\blacksquare$}, and may call APIs from \textcolor{myRed}{$\blacksquare$} and \textcolor{myOrange}{$\blacksquare$}.}
	\label{fig:task_definition}
\end{figure}

Given a code context $\mathbf{x}$, the goal of code generation aims to generate target code $\mathbf{y}$ that solves the programming problem in $\mathbf{x}$.
Figure~\ref{fig:task_definition} presents an example of code context and target code.
In detail, code context includes the natural language description of programming problems in the form of code comments \textcolor{myGray}{$\blacksquare$}, as well as import statements \textcolor{myRed}{$\blacksquare$}, user-defined classes or functions \textcolor{myOrange}{$\blacksquare$}, function headers, and etc.
The entire code generation process can be formalized as $\mathbf{y}=\mathcal{M}(\mathbf{x})$, where $\mathcal{M}$ is code generation model.
In this paper, we focus on private-library-oriented code generation.
A private library, compared to a public one, is characterized by being non-public, i.e., the code library is unseen by $\mathcal{M}$ during pre-training.

\begin{figure}
	\centering
	\includegraphics[scale=0.5]{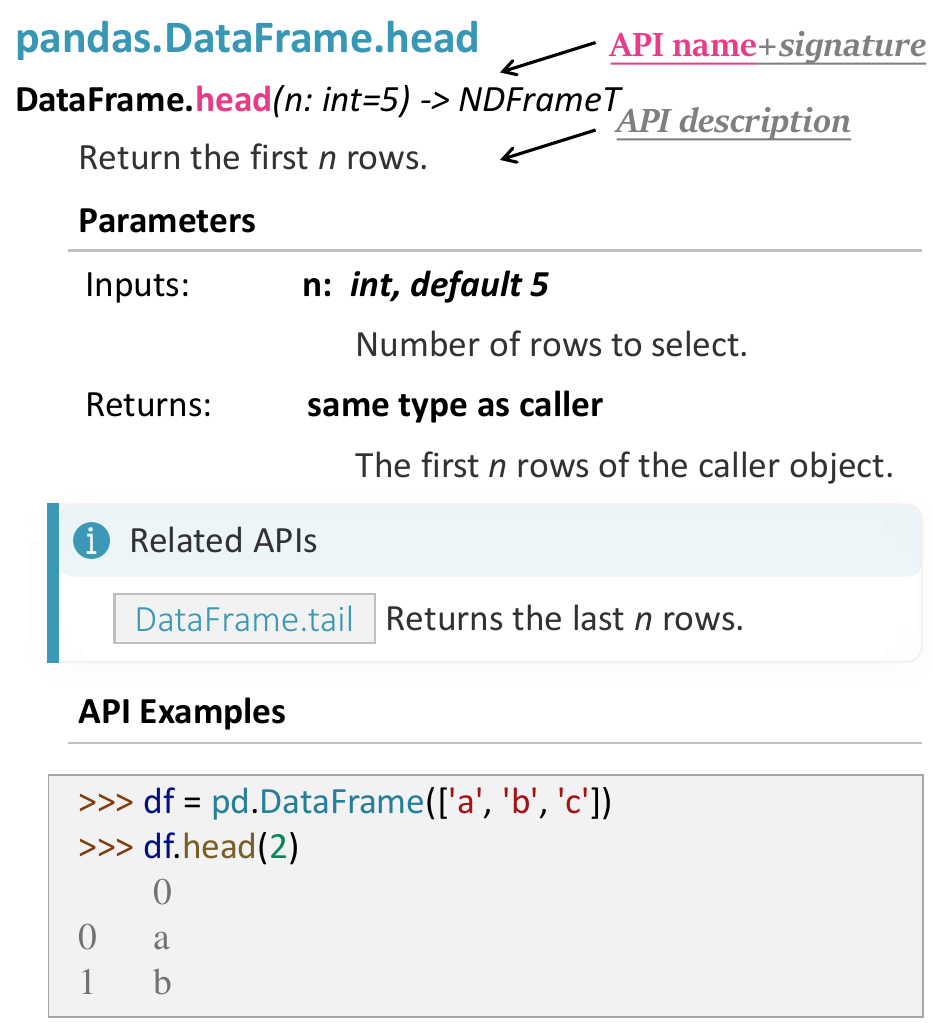}
        \caption{An API instance of \apidoc. It showcases the primary components of each API in \apidoc: API name, signature, description, parameters, related APIs, and API examples.}
	\label{fig:api_documentation}
\end{figure}

\subsection{\apidoc} \label{sec:api_documentation}
\apidoc, also known as API reference, is a comprehensive description of the programming interface that a software library provides to developers.
\apidoc serves as a crucial technical resource to developers for private-library-oriented code generation, enabling them to understand how to use the library and take full advantage of its capabilities.
Considering the significance of \apidoc, we would like to provide a detailed breakdown of its components as illustrated in Figure~\ref{fig:api_documentation}:
\begin{itemize}
    \item \textbf{API Name}: This is a descriptive label used to identify the API and convey its functionality;
    \item \textbf{API Signature}: This outlines the input and output parameters of the API, including data types and expected format;
    \item \textbf{API Description}: This offers a comprehensive explanation of the API's functionality and purpose;
    \item \textbf{API Parameters}: These define the input and return parameters, encompassing information such as data type, default argument values, and descriptions of the parameters;
    \item \textbf{Related APIs}: These are APIs that are closely related to the current API, either in terms of functionality or purpose, and offer similar or complementary capabilities.
    \item \textbf{API Examples}: These provide a working example of how to invoke the API in practice.
\end{itemize}

In this paper, we define \textit{API Basic} as the amalgamation of API Name, API Signature, and API Description.

\section{Framework}

\begin{figure}
	\centering
	\includegraphics[scale=.48]{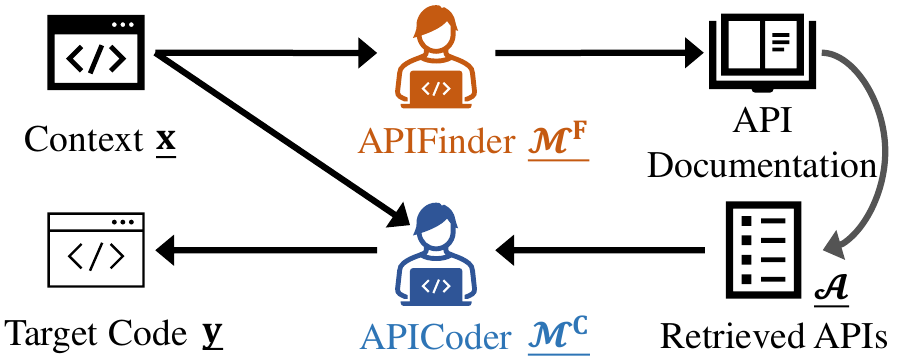}
        \caption{Schematic diagram of our framework: \apifinder first retrieves potentially useful APIs from \apidoc, and then \apicoder generates the target code based on the retrieval APIs.}
	\label{fig:overview_framework}
\end{figure}

Drawing inspiration from how programmers leverage \apidoc to tackle private library scenarios, our framework is devised.
Programmers, as a common practice, first find and locate suitable APIs in \apidoc based on their programming problems, then learn to invoke these APIs to resolve them correctly.
Analogously, as depicted in Figure~\ref{fig:overview_framework}, our framework also consists of two modules: \apifinder $\mathcal{M}^{\text{F}}$, which finds feasible APIs $\mathcal{A}$, and \apicoder $\mathcal{M}^{\text{C}}$, which invokes these APIs in an appropriate manner.
The procedure can be expressed as:
\begin{equation}
    \hspace{30pt} \mathcal{A} = \mathcal{M}^{\text{F}}(\mathbf{x});
    \mathbf{y} = \mathcal{M}^{\text{C}}(\mathcal{A};\mathbf{x}),
\end{equation}
where $\mathbf{x}$ and $\mathbf{y}$ denote code context and target code, $\mathcal{A} \ni \mathbf{a}$ refers to information of all retrieved APIs.
In this study, one of the critical contributions is to investigate which components in \apidoc are beneficial for current code generation models and how to utilize them better.
Therefore, the information of an API $\mathbf{a}$ refers to one or more components in Figure~\ref{fig:api_documentation}.
In essence, our proposed framework solves private library scenarios by dividing the original one-time code generation into retrieval and generation modules, which effectively empowers code generation models to handle private libraries.

\section{Methodology}
Our proposed framework has presented a simple yet novel idea for empowering language models to generate private code snippets.
In the following, we delve into the practical implementation of the framework, including the preparation of the corpus and the finer design details of both \apifinder and \apicoder.

\subsection{Corpus Preparation}
In order to train a neural model for private library scenarios, an intuitive approach is to collect a considerable body of private library corpora for training.
Evidently, obtaining private library corpora is not feasible.
Therefore, we can only train our model on a large corpus of public libraries in hopes of generalizing them to private ones.

In this study, we determine the $31$ most popular Python code libraries, such as Pandas, NumPy, and PyTorch, based on their popularity in Stack Overflow\footnote{\url{https://stackoverflow.com/tags?tab=popular}}.
The names of these libraries and their corresponding API count details can be found in Table~\ref{tab:public_api_name_num}.
For these libraries, we crawl their \apidoc individually.
Furthermore, we decompose each API in \apidoc into sub-components such as API name, signature, description, and examples.
In addition to the \apidoc of these libraries, we also require a substantial collection of code snippets invoking these libraries.
Thus, we initially gathered a $330$GB code corpus from GitHub, containing $60.62$ million Python files.
After a series of data pre-processing strategies, including selecting Python files related to the $31$ code libraries, de-duplication, and formatting, we ultimately acquired approximately $25$GB of Python files comprising a total of $4.54$ million files, referred to as $\mathcal{P}$.

\begin{table*}[width=2.0\linewidth,pos=h]
\centering
\caption{The $31$ public libraries and their corresponding API counts.}
\label{tab:public_api_name_num}
\resizebox{1.0\linewidth}{!}{
\begin{tabular}{cccccccccccl} 
\toprule
\rowcolor[rgb]{0.843,0.843,0.843} Pandas   & NumPy        & sklearn    & PyTorch & TensorFlow & Django & selenium & Matplotlib   & Flask   & SciPy      & Seaborn & ansible   \\
7,094                                      & 12,085       & 53,166     & 124,902 & 32,116     & 24,375 & 4,842    & 439,913      & 31,867  & 153,359    & 161,477 & 40,839    \\
\rowcolor[rgb]{0.843,0.843,0.843} NLTK     & BeatifulSoup & pygame     & PIL     & jieba      & Gensim & spaCy    & transformers & fairseq & SQLAlchemy & Scrapy  & requests  \\
206,816                                    & 22,519       & 70,396     & 127,212 & 26,620     & 37,331 & 239,945  & 652,913      & 158,721 & 54,765     & 3,537   & 39,333    \\
\rowcolor[rgb]{0.843,0.843,0.843} AllenNLP & datasets     & tokenizers & MXNet   & imageio    & pytest & MetPy    &              &         &            &         &           \\
276,088                                    & 136,843      & 195        & 142,070 & 175,878    & 1,047  & 27,429   &              &         &            &         &           \\
\bottomrule
\end{tabular}
}
\end{table*}

\subsection{\apifinder} \label{sec:apifinder}

Given a programming problem description, \apifinder is responsible for matching and finding potentially usable APIs from private library's \apidoc.

\begin{figure*}
	\centering
	\includegraphics[scale=.5]{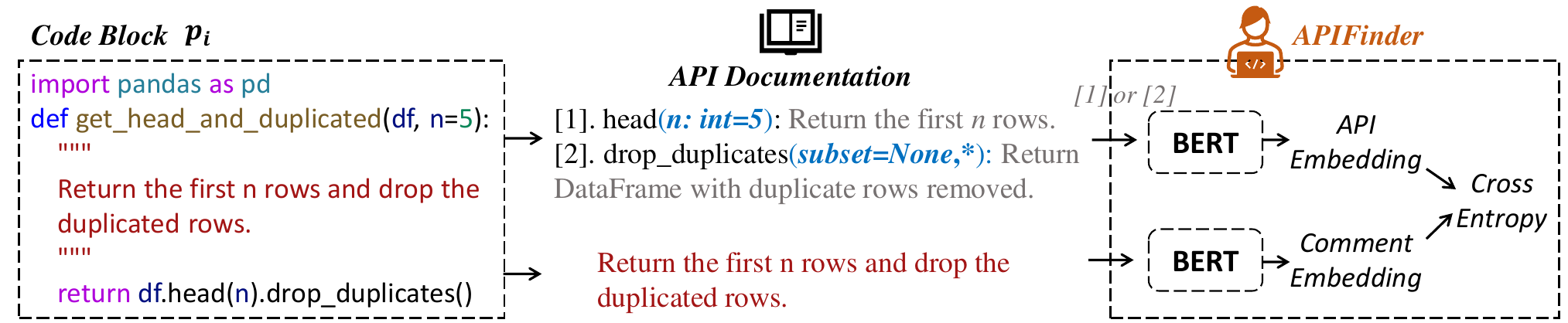}
        \caption{An overview of the preparation of training corpus for \apifinder.}
	\label{fig:apifinder}
\end{figure*}

\paragraph{Training.}
\apifinder can be trained using dense retrieval techniques~\cite{formal2022distillation,rocketqa,santhanam2021colbertv2,xiong2020approximate}.
Dense retrieval techniques mainly consist of two primary approaches: single-encoder and dual-encoder.
Upon receiving a programming problem description, single-encoder models require re-computing the embeddings of each API of \apidoc, while dual-encoder avoids this by pre-computing these APIs offline.
Considering practicality and efficiency during inference, we thus choose to train \apifinder using the dual-encoder one.
To train such a model, a collection of programming problem descriptions and their corresponding APIs is essential.
Thus, we devise a meticulous strategy to extract these data pairs from our collected GitHub corpus $\mathcal{P}$.
Initially, we separate each file $\mathbf{p}$ in $\mathcal{P}$ into $K$ code blocks $(\mathbf{p}_1,\mathbf{p}_2,{\cdots},\mathbf{p}_K)$, where a code block is a well-formed and uninterrupted code snippet such as a method or class. 
This functionality can be automatically implemented via some pip tools, including autopep8, docformatter, and redbaron.
As shown in Figure~\ref{fig:apifinder}, for each code block $\mathbf{p}_i$, we extract its annotation and API names.
For each API name, we locate its corresponding API signature and description by searching the API documentation we crawled.
Note that an API name may match multiple APIs. For instance, the API name \texttt{head} in pandas can correspond to both \texttt{DataFrame.head} and \texttt{Series.head}.
Upon our empirical observation, we conclude that these APIs with the same API name are similar or even identical except for the API path.
Therefore, when encountering an API name corresponding to multiple APIs, we randomly pink one.
For each code block $\mathbf{p}_i$, we have obtained multiple data pairs of programming problem description $\mathbf{d}$ and API information $\mathbf{a}$, where each $\mathbf{a}$ includes its API name, signature, and description\footnote{In all our experiments, we solely use the first sentence of the API description, as it sufficiently summarizes the content.}.
We regard the extracted $\mathbf{d}$ and $\mathbf{a}$ from a code block as a positive sample.
For the negative ones, we randomly select $n$ APIs irrelevant to $\mathbf{d}$, with $n$ set to $8$ in our experiments.
Altogether, we extract $40.37$ million instances $(\mathbf{d},\mathbf{a},\hat{\mathbf{a}}_1,\hat{\mathbf{a}}_2,\dots,\hat{\mathbf{a}}_n)$ to train our \apifinder.
Following RocketQA~\cite{rocketqa}, we encode $\mathbf{d}$ and $\mathbf{a}$ into 768-dimensional embedding separately using two BERT-base models~\cite{bert} named $E_{\mathbf{d}}$ and $E_{\mathbf{a}}$.
We then compute the similarity between these two embeddings using the dot product and optimize it with cross-entropy loss.

\paragraph{Inference.}
After successfully training our \apifinder on a vast array of public libraries, we can leverage it to retrieve private APIs from \apidoc.
In our practical private library scenarios, the \apidoc is typically considered a stable offline resource that remains unchanged except for version updates.
Therefore, $E_{\mathbf{a}}$ can pre-encode all APIs in the \apidoc offline and index them in FAISS~\cite{faiss} for efficient retrieval.
Upon receiving a new programming problem description, we merely compute its embedding using $E_{\mathbf{d}}$, and subsequently retrieve the nearest API(s) from the offline pre-encoded APIs according to spatial proximity.

\begin{figure}
	\centering
	\includegraphics[scale=.5]{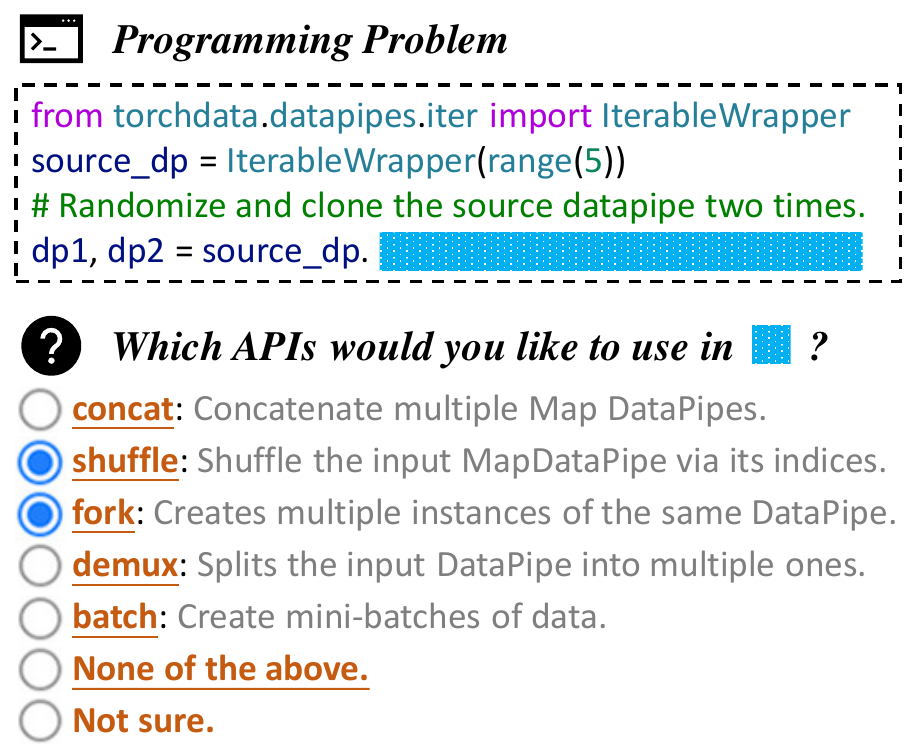}
        \caption{User interaction interface with \apifinder: users can choose one or more APIs from \apifinder's top $5$ recommendations for \apicoder to use.}
	\label{fig:human_in_the_loop}
\end{figure}

\paragraph{User's Optional Engagement with \apifinder.}
These APIs retrieved by \apifinder can be directly input to \apicoder for generating code that invokes private libraries.
To provide \apicoder with more accurate APIs, we design a pipeline that allows user engagement with \apifinder.
Specifically, as shown in Figure~\ref{fig:human_in_the_loop}, we furnish users with an interactive interface that affords them to select one or more from the top $5$ APIs retrieved by \apifinder.
We also provide "None of the above" and "Not sure" as options for the user if they find the APIFinder retrieval results lacking a definitive answer or uncertain.
If "None of the above" is selected, \apicoder will prompt no APIs. If "Not sure" is chosen, the first two retrieved APIs will be prompted to \apicoder by default.
In our designed interface, we present only the API name and the first sentence of its description, as excessive information may result in reader fatigue and a negative impact.

\subsection{\apicoder} \label{sec:apicoder}

\apicoder leverages the APIs retrieved by \apifinder to write private code.
Prior to invoking these APIs, programmers are required to learn them.
Similarly, large language models also require incorporating these APIs into the code context before invocation via a well-designed prompt.

\paragraph{Off-the-shelf \apicoder.}
Retrieval-based generation facilitates the powerful generative capability of new samples in LLMs via the provision of context-rich guidance.
Therefore, technically speaking, existing code generation models, such as Codex~\cite{codex}, InCoder~\cite{incoder}, and \codegen~\cite{codegen}, can be used directly as our \apicoder by providing rich private API information.

\begin{figure*}
	\centering
	\includegraphics[scale=.47]{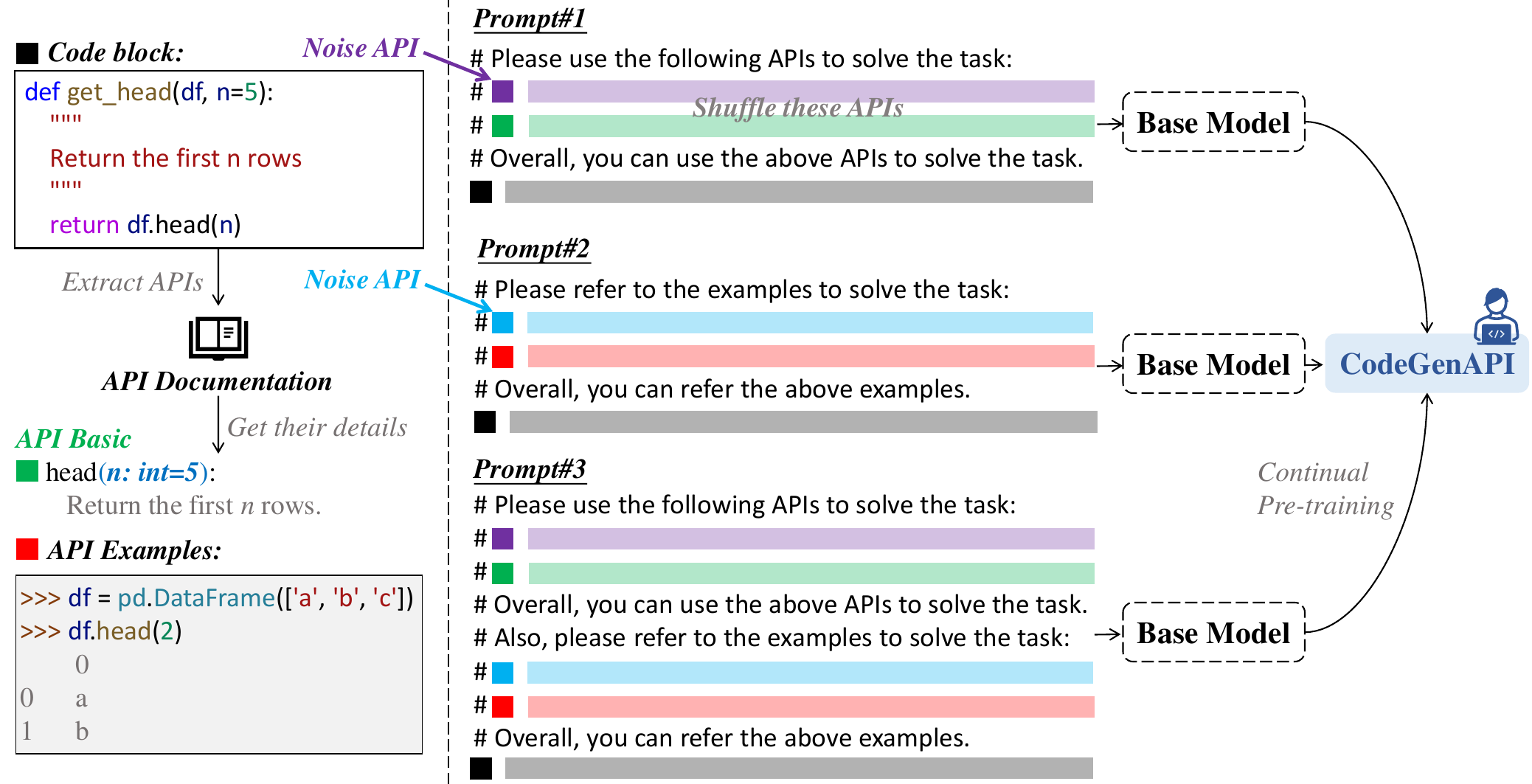}
        \caption{An overview of the preparation of training corpus for \codegenapi. For visual simplicity, we use colors instead of text of the left to illustrate how to design prompts of the right. We utilize \textcolor{myPurple}{$\blacksquare$} and \textcolor{myBlue}{$\blacksquare$} to represent the noise API, with their format being consistently \textcolor{myGreen}{$\blacksquare$} and \textcolor{myRed}{$\blacksquare$}, respectively.}
	\label{fig:apicoder}
\end{figure*}

\paragraph{Training an Advanced \apicoder.}
Off-the-shelf code generation models can leverage retrieval-based generation to invoke private APIs, yet there is significant potential for improvement as they have not been explicitly trained on how to invoke APIs.
Thus, to further enhance the code generation performance of private libraries, we propose an intriguing idea to continually train an advanced \apicoder.
To bring this idea to fruition, we require a substantial training corpus comprising code blocks along with their corresponding API information.
However, the existing GitHub corpus often lacks API information prior to each code block.
Consequently, we add relevant API information preceding each code block to continually pre-train off-the-shelf models.
In our experiments, we use \codegen~\cite{codegen} as our base model as it is widely used among publicly available models. We name the trained model as \codegenapi.
For the training corpus, we first segment each Python file $\mathbf{p}\in\mathcal{P}$ into $K$ code blocks $(\mathbf{p}_1,\mathbf{p}_2,{\cdots},\mathbf{p}_K)$, as done in \apifinder.
Then, we extract relevant APIs $\mathcal{A}_i$ for each code block $\mathbf{p}_i$.
Finally, we concatenate the extracted APIs to the front of each code block to obtain the corpus $(\mathcal{A}_1,\mathbf{p}_1,\mathcal{A}_2,\mathbf{p}_2,{\cdots},\mathcal{A}_K,\mathbf{p}_K)$.
As stated in Section~\ref{sec:api_documentation}, API documentation includes various components such as API Basics, Examples, and Parameters.
Placing all components of each API in front of the code block is impractical due to excessive length.
Thus, as shown in Figure~\ref{fig:apicoder}, we design three prompts to train \codegenapi separately: API Basic, API Examples, and both combined.
Moreover, to improve the robustness of \codegenapi, we incorporate noise APIs with a probability of $5$\% and shuffle these APIs in $\mathcal{A}$, as the APIs returned by \apifinder may contain incorrect APIs and are unordered.
During the training of \codegenapi, we employ a re-sampling strategy.
This strategy prioritizes high-quality Python files in sampling and reduces the selection of low-quality files.
Specifically, the re-sampling weight ${w}$ of each Python file can be defined as:
\begin{equation}
    \begin{split}
        &{w} = { w}_{\rm api}\times{w}_{\rm star}\times{ w}_{\rm ut}, \\
        &{w}_{\rm api} = 5.0 - \log(\frac{M_{\rm api}}{N_{\rm api}}).{\rm clip}(0,5)\times0.2, \\
        &{w}_{\rm star} = 1.0 + \log(N_{\rm star}+1).{\rm clip}(0,5)\times0.2, \\
        &{w}_{\rm ut} = (0.5 + (1 - {R}_{{\rm ut}})).{\rm clip}(0,1),
    \end{split}
    \label{equation:resample}
\end{equation}
where $.\rm clip$$(x,y)$ restricts the value to a range $[x,y]$, 
$N_{\rm api}$ denotes the count of API names in this file, 
$M_{\rm api}$ represents the count of APIs in cases where a single API name matches multiple APIs,
$N_{\rm star}$ signifies the number of stars for the repository,
and $R_{\rm ut}$ is the unit test function rate, calculated as the division of unit test functions by the total functions.

\paragraph{Inference.}
Both off-the-shelf and our advanced \apicoder are Transformer-based generative language models.
During inference, our \apicoder predicts the probability distribution of the next token based on the given code context until encountering one of the following pre-defined stop markers: ``\texttt{\textbackslash nclass}'', ``\texttt{\textbackslash ndef}'', 
``\texttt{\textbackslash nprint}'', ``\texttt{\textbackslash n\#}'', or
``\texttt{\textbackslash nif}''.
After predicting the probability distribution, we require the decoding of these probabilities.
Currently, generative language models offer a variety of decoding techniques, such as temperature decoding, greedy decoding, and nucleus decoding.
In code generation scenarios, temperature decoding is widely adopted~\cite{codex,alphacode} because it allows for a suitable balance between diversity and quality in generated code, while other decoding techniques cannot.
Consequently, all of our experiments use temperature decoding for inference.

\section{Benchmark Construction} \label{sec:benchmark_construction}
In software development, the use of private libraries is a prevalent practice. However, as of yet, no benchmarks for private libraries have been crafted.
A primary reason for this is that crafting a comprehensive benchmark for private libraries poses many considerable challenges.
Specifically, these challenges are manifested in the following aspects: (1) ensuring that LLMs have not seen the library during training, (2) providing high-quality implementation code and \apidoc, (3) offering programming problems of varying difficulties, and (4) annotating comprehensive test cases for each programming problem.
To tackle the above challenges, we release four benchmarks for private libraries, namely \torchdataeval, \torchdatacomplexeval, \monkeyeval, and \beatnumeval.

\paragraph{\torchdataeval.}
After surveying numerous code libraries, we chose TorchData\footnote{\url{https://pytorch.org/data/beta/index.html}}, released in May $2022$, as our private library for its high-quality implementation code and comprehensive \apidoc. More importantly, all models in our experiments, including Codex and \codegen, are pre-trained on the GitHub corpus prior to the aforementioned release date, ensuring that the LLMs have no prior exposure to the library.
Before crafting programming problems, we thoroughly studied the \apidoc and hands-on each API to guarantee our proficiency with the library.
Subsequently, we manually crafted $50$ programming problems and provided comprehensive test cases for each.
Additionally, we invited two volunteers with over $3$ years of Python coding experience to review them to ensure their soundness and correctness.
In fact, our worry is that the complexity of the programming challenges may surpass the capability of the LLM.
Therefore, we control the difficulty level of these programming problems to be relatively simple. Concretely, programming problems that include $1$ or $2$ APIs make up approximately $90\%$.

\paragraph{\torchdatacomplexeval.}
Astoundingly, our experiments reveal that LLMs can perform relatively well on \torchdataeval, indicating their capability to generate some simple private code snippets.
One research question thus arises: can LLMs cope with more complex private libraries? 
Therefore, we developed a more complex benchmark base on TorchData, consisting of $50$ programming problems, named \torchdatacomplexeval.
Unlike \torchdataeval, which manually creates programming problems by referencing examples in API documentation, \torchdatacomplexeval directly adapts practical projects on GitHub.
Specifically, we carefully convert real-world projects into executable and well-annotated, also equipped with comprehensive test cases.
Due to the intricacy of these projects, \torchdatacomplexeval's programming problems typically involve a high volume of APIs, often exceeding $10$.
TorchData is a library for building data pipelines that can handle diverse data modalities. Considering the potential bias among these modalities, \torchdatacomplexeval encompasses common ones like text, vision, and audio.
We also invited two volunteers to review these programming problems, as done in crafting \torchdataeval.

\paragraph{\monkeyeval\&\beatnumeval.}
As discussed above, a qualified private library possesses rigorous criteria.
So, discovering a suitable private library like TorchData proves to be exceptionally challenging.
Hence, besides \torchdataeval and \torchdatacomplexeval, we created two pseudo-private library benchmarks, named \monkeyeval and \beatnumeval, by adapting two off-the-shelf public ones~\cite{CERT}, named \pandaseval and \numpyeval, each with $101$ programming problems.
We manually paraphrased the keywords in \pandaseval and \numpyeval.
As depicted in Figure~\ref{fig:example_llm_private}, we paraphrased ``\texttt{pandas}'' to ``\texttt{monkey}'', ``\texttt{dataframe}'' to ``\texttt{knowledgeframe}'', and ``\texttt{isin}'' to ``\texttt{iscontain}''.
Further details on the paraphrasing are provided in Appendix~\ref{apx:keyword_conversion}.
Besides the keywords, we also thoroughly rephrased their API documentation to ensure that LLMs have not seen them.

\begin{table*}[width=2.0\linewidth,pos=h]
\centering
\caption{\Passk (\%) of $17$ code generation models on four private library benchmarks. 
No API, Oracle, TopN, and Human denote using no API, perfect APIs, APIs retrieved from \apifinder, and APIs selected from Top5 by humans as extra prompt. B, E, and BE denote using API's basic information (API name, signature, description), examples, and both as extra prompt. Values in \textcolor{red}{red}/\textcolor{darkgreen}{green} represent improvement/decline relative to the No API setting.}
\label{tab:main_results}
\resizebox{\linewidth}{!}{
\begin{tabular}{ll|ll:ll:ll:ll} 
\toprule
\multirow{2}{*}{\textbf{\apicoder }} & \multicolumn{1}{l}{\multirow{2}{*}{\textbf{APIFinder }}} & \multicolumn{2}{c}{\textbf{TorchDataEval}}                                                                                                                    & \multicolumn{2}{c}{\begin{tabular}[c]{@{}c@{}}\textbf{TorchData}\\\textbf{ComplexEval}\end{tabular}}                                                         & \multicolumn{2}{c}{\textbf{MonkeyEval}}                                                                                                                                 & \multicolumn{2}{c}{\textbf{BeatNumEval}}                                                                                                                                 \\ 
\cline{3-10}
                                     & \multicolumn{1}{l}{}                                     & \multicolumn{1}{c}{pass@1}                                                    & \multicolumn{1}{c}{pass@10}                                                   & \multicolumn{1}{c}{pass@1}                                                   & \multicolumn{1}{c}{pass@10}                                                   & \multicolumn{1}{c}{pass@1}                                                    & \multicolumn{1}{c}{pass@10}                                                             & \multicolumn{1}{c}{pass@1}                                                    & \multicolumn{1}{c}{pass@10}                                                              \\ 
\hline\hline
CodeGPT 124M                         & Top2/B                                                   & 0.48                                                                          & 2.83                                                                          & 0.00                                                                         & 0.00                                                                          & 0.18                                                                          & 1.02                                                                                    & 0.12                                                                          & 0.96                                                                                     \\
GPT-CC 125M                          & Top2/B                                                   & 0.00                                                                          & 0.00                                                                          & 0.00                                                                         & 0.00                                                                          & 0.04                                                                          & 0.34                                                                                    & 0.12                                                                          & 0.92                                                                                     \\
GPT-CC 1.3B                          & Top2/B                                                   & 0.00                                                                          & 0.00                                                                          & 0.20                                                                         & 1.34                                                                          & 0.00                                                                          & 0.00                                                                                    & 0.10                                                                          & 0.66                                                                                     \\
CodeParrot 110M                      & Top2/B                                                   & 1.96                                                                          & 9.32                                                                          & 0.08                                                                         & 0.76                                                                          & 0.55                                                                          & 1.84                                                                                    & 2.69                                                                          & 10.56                                                                                    \\
CodeParrot 1.5B                      & Top2/B                                                   & 4.00                                                                          & 12.25                                                                         & 0.00                                                                         & 0.00                                                                          & 1.58                                                                          & 3.97                                                                                    & 2.48                                                                          & 6.07                                                                                     \\
PyCodeGPT 110M                       & Top2/B                                                   & 4.36                                                                          & 16.66                                                                         & 2.12                                                                         & 6.19                                                                          & 2.26                                                                          & 9.10                                                                                    & 5.80                                                                          & 19.17                                                                                    \\
CodeT5 770M                          & Top2/B                                                   & 16.62                                                                         & 33.47                                                                         & 3.60                                                                         & 5.95                                                                          & 4.16                                                                          & 14.88                                                                                   & 9.60                                                                          & 24.23                                                                                    \\
PolyCoder 160M                       & Top2/B                                                   & 3.38                                                                          & 9.43                                                                          & 0.00                                                                         & 0.00                                                                          & 0.83                                                                          & 4.16                                                                                    & 2.90                                                                          & 12.25                                                                                    \\
PolyCoder 400M                       & Top2/B                                                   & 3.60                                                                          & 12.57                                                                         & 0.00                                                                         & 0.00                                                                          & 0.99                                                                          & 3.70                                                                                    & 4.70                                                                          & 16.34                                                                                    \\
PolyCoder 2.7B                       & Top2/B                                                   & 5.60                                                                          & 14.28                                                                         & 0.40                                                                         & 2.68                                                                          & 1.68                                                                          & 6.42                                                                                    & 5.05                                                                          & 15.69                                                                                    \\
InCoder 1.3B                         & Top2/B                                                   & 6.00                                                                          & 15.59                                                                         & 0.80                                                                         & 3.62                                                                          & 3.56                                                                          & 13.93                                                                                   & 5.35                                                                          & 16.66                                                                                    \\
InCoder 6.7B                         & Top2/B                                                   & 7.00                                                                          & 22.63                                                                         & 2.80                                                                         & 4.00                                                                          & 5.25                                                                          & 18.49                                                                                   & 8.42                                                                          & 27.90                                                                                    \\
SantaCoder 1.1B                      & Top2/B                                                   & 19.20                                                                         & 32.62                                                                         & 2.40                                                                         & 4.00                                                                          & 4.16                                                                          & 12.91                                                                                   & 11.68                                                                         & 22.77                                                                                    \\ 
\hline
\multirow{11}{*}{\codegen 350M}       & No API                                                   & 4.40                                                                          & 17.76                                                                         & 1.12                                                                         & 5.79                                                                          & 2.38                                                                          & 9.77                                                                                    & 6.24                                                                          & 24.58                                                                                    \\
                                     & {\cellcolor[rgb]{0.792,0.918,1}}Oracle/B                 & {\cellcolor[rgb]{0.792,0.918,1}}7.22\textcolor{red}{\textsuperscript{2.82}}   & {\cellcolor[rgb]{0.792,0.918,1}}28.53\textcolor{red}{\textsuperscript{10.77}} & {\cellcolor[rgb]{0.792,0.918,1}}1.36\textcolor{red}{\textsuperscript{0.24}}  & {\cellcolor[rgb]{0.792,0.918,1}}6.52\textsuperscript{\textcolor{red}{0.73}}   & {\cellcolor[rgb]{0.792,0.918,1}}3.29\textcolor{red}{\textsuperscript{0.91}}   & {\cellcolor[rgb]{0.792,0.918,1}}11.38\textsuperscript{\textcolor{red}{1.61}}            & {\cellcolor[rgb]{0.792,0.918,1}}11.19\textsuperscript{\textcolor{red}{4.95}}  & {\cellcolor[rgb]{0.792,0.918,1}}28.92\textsuperscript{\textcolor{red}{4.34}}             \\
                                     & {\cellcolor[rgb]{0.792,0.918,1}}Oracle/E                 & {\cellcolor[rgb]{0.792,0.918,1}}13.12\textcolor{red}{\textsuperscript{8.72}}  & {\cellcolor[rgb]{0.792,0.918,1}}30.84\textsuperscript{\textcolor{red}{13.08}} & {\cellcolor[rgb]{0.792,0.918,1}}1.92\textsuperscript{\textcolor{red}{0.80}}  & {\cellcolor[rgb]{0.792,0.918,1}}6.71\textsuperscript{\textcolor{red}{0.92}}   & {\cellcolor[rgb]{0.792,0.918,1}}3.41\textsuperscript{\textcolor{red}{1.03}}   & {\cellcolor[rgb]{0.792,0.918,1}}15.41\textsuperscript{\textcolor{red}{5.64}}            & {\cellcolor[rgb]{0.792,0.918,1}}6.53\textsuperscript{\textcolor{red}{0.29}}   & {\cellcolor[rgb]{0.792,0.918,1}}23.18\textsuperscript{\textcolor[rgb]{0,0.502,0}{0.78}}  \\
                                     & {\cellcolor[rgb]{0.792,0.918,1}}Oracle/BE                & {\cellcolor[rgb]{0.792,0.918,1}}12.06\textsuperscript{\textcolor{red}{7.66}}  & {\cellcolor[rgb]{0.792,0.918,1}}31.66\textsuperscript{\textcolor{red}{13.90}} & {\cellcolor[rgb]{0.792,0.918,1}}1.76\textsuperscript{\textcolor{red}{0.64}}  & {\cellcolor[rgb]{0.792,0.918,1}}8.62\textsuperscript{\textcolor{red}{2.83}}   & {\cellcolor[rgb]{0.792,0.918,1}}2.65\textsuperscript{\textcolor{red}{0.27}}   & {\cellcolor[rgb]{0.792,0.918,1}}11.61\textsuperscript{\textcolor{red}{1.84}}            & {\cellcolor[rgb]{0.792,0.918,1}}7.52\textsuperscript{\textcolor{red}{1.28}}   & {\cellcolor[rgb]{0.792,0.918,1}}24.81\textsuperscript{\textcolor{red}{0.23}}             \\
                                     & Top1/B                                                   & 6.04\textsuperscript{\textcolor{red}{1.64}}                                   & 21.73\textsuperscript{\textcolor{red}{3.97}}                                  & 1.28\textsuperscript{\textcolor{red}{0.16}}                                  & 6.21\textsuperscript{\textcolor{red}{0.42}}                                   & 3.37\textcolor{red}{\textsuperscript{0.99}}                                   & 11.02\textsuperscript{\textcolor{red}{1.25}}                                            & 8.87\textsuperscript{\textcolor{red}{2.63}}                                   & 24.28\textsuperscript{\textcolor[rgb]{0,0.502,0}{0.30}}                                  \\
                                     & Top2/B                                                   & 5.72\textsuperscript{\textcolor{red}{1.32}}                                   & 20.36\textsuperscript{2.60}                                                   & 1.36\textsuperscript{\textcolor{red}{0.24}}                                  & 7.20\textsuperscript{\textcolor{red}{1.41}}                                   & 3.76\textsuperscript{\textcolor{red}{1.38}}                                   & 12.39\textsuperscript{\textcolor{red}{2.62}}                                            & 9.31\textsuperscript{\textcolor{red}{3.07}}                                   & 24.33\textsuperscript{\textcolor[rgb]{0,0.502,0}{0.25}}                                  \\
                                     & Top2/E                                                   & 6.66\textsuperscript{\textcolor{red}{2.26}}                                   & 16.39\textsuperscript{\textcolor[rgb]{0,0.502,0}{1.37}}                       & 1.12\textsuperscript{\textcolor{red}{0.00}}                                  & 6.75\textsuperscript{\textcolor{red}{0.96}}                                   & 2.42\textsuperscript{\textcolor{red}{0.04}}                                   & 10.84\textsuperscript{\textcolor{red}{1.07}}                                            & 5.50\textsuperscript{\textcolor[rgb]{0,0.502,0}{0.74}}                        & 16.37\textsuperscript{\textcolor[rgb]{0,0.502,0}{8.21}}                                  \\
                                     & Top2/BE                                                  & 5.44\textsuperscript{\textcolor{red}{1.04}}                                   & 18.10\textsuperscript{\textcolor{red}{0.34}}                                  & 1.44\textsuperscript{\textcolor{red}{0.32}}                                  & 7.66\textsuperscript{\textcolor{red}{1.87}}                                   & 1.99\textsuperscript{\textcolor[rgb]{0,0.502,0}{0.39}}                        & 9.22\textsuperscript{\textcolor[rgb]{0,0.502,0}{0.55}}                                  & 4.28\textsuperscript{\textcolor[rgb]{0,0.502,0}{1.96}}                        & 14.73\textsuperscript{\textcolor[rgb]{0,0.502,0}{9.85}}                                  \\
                                     & Top3/B                                                   & 6.28\textsuperscript{\textcolor{red}{1.88}}                                   & 22.06\textsuperscript{\textcolor{red}{4.30}}                                  & 1.20\textsuperscript{\textcolor{red}{0.08}}                                  & 7.49\textsuperscript{\textcolor{red}{1.70}}                                   & 3.84\textsuperscript{\textcolor{red}{1.46}}                                   & 11.61\textsuperscript{\textcolor{red}{1.84}}                                            & 9.07\textsuperscript{\textcolor{red}{2.83}}                                   & 25.82\textsuperscript{\textcolor{red}{1.24}}                                             \\
                                     & Top5/B                                                   & 7.04\textsuperscript{\textcolor{red}{2.64}}                                   & 25.01\textsuperscript{\textcolor{red}{7.25}}                                  & 1.12\textsuperscript{\textcolor{red}{0.00}}                                  & 6.63\textsuperscript{\textcolor{red}{0.84}}                                   & 4.08\textsuperscript{\textcolor{red}{1.70}}                                   & 12.74\textsuperscript{\textcolor{red}{2.97}}                                            & 8.93\textsuperscript{\textcolor{red}{2.69}}                                   & 26.26\textsuperscript{\textcolor{red}{1.68}}                                             \\
                                     & {\cellcolor[rgb]{1,0.941,0.835}}Human/B                  & {\cellcolor[rgb]{1,0.941,0.835}}6.62\textsuperscript{\textcolor{red}{2.22}}   & {\cellcolor[rgb]{1,0.941,0.835}}26.03\textsuperscript{\textcolor{red}{8.27}}  & {\cellcolor[rgb]{1,0.941,0.835}}1.44\textsuperscript{\textcolor{red}{0.32}}  & {\cellcolor[rgb]{1,0.941,0.835}}6.81\textsuperscript{\textcolor{red}{1.02}}   & {\cellcolor[rgb]{1,0.941,0.835}}3.35\textsuperscript{\textcolor{red}{0.97}}   & {\cellcolor[rgb]{1,0.941,0.835}}11.78\textsuperscript{\textcolor{red}{2.01}}            & {\cellcolor[rgb]{1,0.941,0.835}}11.39\textsuperscript{\textcolor{red}{5.15}}  & {\cellcolor[rgb]{1,0.941,0.835}}30.12\textsuperscript{\textcolor{red}{5.54}}             \\ 
\hline
\multirow{11}{*}{\codegen 2B~}        & No API                                                   & 8.80                                                                          & 20.92                                                                         & 5.60                                                                         & 13.24                                                                         & 4.65                                                                          & 12.47                                                                                   & 8.71                                                                          & 30.21                                                                                    \\
                                     & {\cellcolor[rgb]{0.792,0.918,1}}Oracle/B                 & {\cellcolor[rgb]{0.792,0.918,1}}18.80\textsuperscript{\textcolor{red}{10.00}} & {\cellcolor[rgb]{0.792,0.918,1}}42.93\textsuperscript{\textcolor{red}{22.01}} & {\cellcolor[rgb]{0.792,0.918,1}}5.60\textsuperscript{\textcolor{red}{0.00}}  & {\cellcolor[rgb]{0.792,0.918,1}}13.59\textsuperscript{\textcolor{red}{0.35}}  & {\cellcolor[rgb]{0.792,0.918,1}}7.23\textsuperscript{\textcolor{red}{2.58}}   & {\cellcolor[rgb]{0.792,0.918,1}}17.25\textsuperscript{\textcolor{red}{4.78}}            & {\cellcolor[rgb]{0.792,0.918,1}}16.73\textsuperscript{\textcolor{red}{8.02}}  & {\cellcolor[rgb]{0.792,0.918,1}}37.79\textsuperscript{\textcolor{red}{7.58}}             \\
                                     & {\cellcolor[rgb]{0.792,0.918,1}}Oracle/E                 & {\cellcolor[rgb]{0.792,0.918,1}}23.40\textsuperscript{\textcolor{red}{14.60}} & {\cellcolor[rgb]{0.792,0.918,1}}41.96\textsuperscript{\textcolor{red}{21.04}} & {\cellcolor[rgb]{0.792,0.918,1}}7.00\textsuperscript{\textcolor{red}{1.40}}  & {\cellcolor[rgb]{0.792,0.918,1}}18.49\textsuperscript{\textcolor{red}{5.25}}  & {\cellcolor[rgb]{0.792,0.918,1}}5.74\textsuperscript{\textcolor{red}{1.09}}   & {\cellcolor[rgb]{0.792,0.918,1}}14.08\textsuperscript{\textcolor{red}{1.61}}            & {\cellcolor[rgb]{0.792,0.918,1}}10.69\textsuperscript{\textcolor{red}{1.98}}  & {\cellcolor[rgb]{0.792,0.918,1}}26.46\textsuperscript{\textcolor[rgb]{0,0.502,0}{3.75}}  \\
                                     & {\cellcolor[rgb]{0.792,0.918,1}}Oracle/BE                & {\cellcolor[rgb]{0.792,0.918,1}}24.00\textsuperscript{\textcolor{red}{15.20}} & {\cellcolor[rgb]{0.792,0.918,1}}43.55\textsuperscript{\textcolor{red}{22.63}} & {\cellcolor[rgb]{0.792,0.918,1}}6.40\textsuperscript{\textcolor{red}{0.80}}  & {\cellcolor[rgb]{0.792,0.918,1}}18.59\textsuperscript{\textcolor{red}{5.35}}  & {\cellcolor[rgb]{0.792,0.918,1}}5.74\textsuperscript{\textcolor{red}{1.09}}   & {\cellcolor[rgb]{0.792,0.918,1}}11.49\textsuperscript{\textcolor[rgb]{0,0.502,0}{0.98}} & {\cellcolor[rgb]{0.792,0.918,1}}9.50\textsuperscript{\textcolor{red}{0.79}}   & {\cellcolor[rgb]{0.792,0.918,1}}30.82\textsuperscript{\textcolor{red}{0.61}}             \\
                                     & Top1/B                                                   & 10.20\textsuperscript{\textcolor{red}{1.40}}                                  & 27.98\textsuperscript{\textcolor{red}{7.06}}                                  & 5.60\textsuperscript{\textcolor{red}{0.00}}                                  & 13.61\textsuperscript{\textcolor{red}{0.37}}                                  & 6.93\textsuperscript{\textcolor{red}{2.28}}                                   & 17.87\textsuperscript{\textcolor{red}{5.40}}                                            & 13.07\textsuperscript{\textcolor{red}{4.36}}                                  & 31.58\textsuperscript{\textcolor{red}{1.37}}                                             \\
                                     & Top2/B                                                   & 13.00\textsuperscript{\textcolor{red}{4.20}}                                  & 35.04\textsuperscript{\textcolor{red}{14.12}}                                 & 5.80\textsuperscript{\textcolor{red}{0.2}}                                   & 13.78\textsuperscript{\textcolor{red}{0.54}}                                  & 8.42\textsuperscript{\textcolor{red}{3.77}}                                   & 20.12\textsuperscript{\textcolor{red}{7.65}}                                            & 12.97\textsuperscript{\textcolor{red}{4.26}}                                  & 30.82\textsuperscript{\textcolor{red}{0.61}}                                             \\
                                     & Top2/E                                                   & 17.00\textsuperscript{\textcolor{red}{8.20}}                                  & 30.74\textsuperscript{\textcolor{red}{9.82}}                                  & 5.80\textsuperscript{\textcolor{red}{0.2}}                                   & 13.23\textsuperscript{\textcolor[rgb]{0,0.502,0}{0.01}}                       & 4.55\textsuperscript{\textcolor[rgb]{0,0.502,0}{0.10}}                        & 12.92\textsuperscript{\textcolor{red}{0.45}}                                            & 7.52\textsuperscript{\textcolor[rgb]{0,0.502,0}{1.19}}                        & 18.03\textsuperscript{\textcolor[rgb]{0,0.502,0}{12.18}}                                 \\
                                     & Top2/BE                                                  & 16.00\textsuperscript{\textcolor{red}{7.20}}                                  & 31.17\textsuperscript{\textcolor{red}{10.25}}                                 & 6.80\textsuperscript{\textcolor{red}{1.2}}                                   & 17.21\textsuperscript{\textcolor{red}{3.97}}                                  & 4.95\textsuperscript{\textcolor{red}{0.30}}                                   & 12.49\textsuperscript{\textcolor{red}{0.02}}                                            & 3.96\textsuperscript{\textcolor[rgb]{0,0.502,0}{4.75}}                        & 10.22\textsuperscript{\textcolor[rgb]{0,0.502,0}{19.99}}                                 \\
                                     & Top3/B                                                   & 12.60\textsuperscript{\textcolor{red}{3.80}}                                  & 36.19\textsuperscript{\textcolor{red}{15.27}}                                 & 4.40\textsuperscript{\textcolor[rgb]{0,0.502,0}{1.2}}                        & 12.61\textsuperscript{\textcolor[rgb]{0,0.502,0}{0.63}}                       & 8.12\textsuperscript{\textcolor{red}{3.47}}                                   & 18.58\textsuperscript{\textcolor{red}{6.11}}                                            & 11.58\textsuperscript{\textcolor{red}{2.87}}                                  & 30.29\textsuperscript{\textcolor{red}{0.08}}                                             \\
                                     & Top5/B                                                   & 10.60\textsuperscript{\textcolor{red}{1.80}}                                  & 33.95\textsuperscript{\textcolor{red}{13.03}}                                 & 5.20\textsuperscript{\textcolor[rgb]{0,0.502,0}{0.4}}                        & 13.60\textsuperscript{\textcolor{red}{0.36}}                                  & 9.41\textsuperscript{\textcolor{red}{4.76}}                                   & 20.23\textsuperscript{\textcolor{red}{7.76}}                                            & 11.78\textsuperscript{\textcolor{red}{3.07}}                                  & 32.16\textsuperscript{\textcolor{red}{1.95}}                                             \\
                                     & {\cellcolor[rgb]{1,0.941,0.835}}Human/B                  & {\cellcolor[rgb]{1,0.941,0.835}}12.20\textsuperscript{\textcolor{red}{3.40}}  & {\cellcolor[rgb]{1,0.941,0.835}}29.51\textsuperscript{\textcolor{red}{8.59}}  & {\cellcolor[rgb]{1,0.941,0.835}}5.88\textsuperscript{\textcolor{red}{0.28}}  & {\cellcolor[rgb]{1,0.941,0.835}}13.60\textsuperscript{\textcolor{red}{0.36}}  & {\cellcolor[rgb]{1,0.941,0.835}}7.13\textsuperscript{\textcolor{red}{2.48}}   & {\cellcolor[rgb]{1,0.941,0.835}}17.01\textsuperscript{\textcolor{red}{4.54}}            & {\cellcolor[rgb]{1,0.941,0.835}}16.24\textsuperscript{\textcolor{red}{7.53}}  & {\cellcolor[rgb]{1,0.941,0.835}}35.77\textsuperscript{\textcolor{red}{5.56}}             \\ 
\hline
\multirow{11}{*}{\codegen 6B}         & No API                                                   & 9.28                                                                          & 27.63                                                                         & 6.40                                                                         & 13.68                                                                         & 5.26                                                                          & 14.59                                                                                   & 12.58                                                                         & 33.80                                                                                    \\
                                     & {\cellcolor[rgb]{0.792,0.918,1}}Oracle/B                 & {\cellcolor[rgb]{0.792,0.918,1}}24.72\textsuperscript{\textcolor{red}{15.44}} & {\cellcolor[rgb]{0.792,0.918,1}}47.32\textsuperscript{\textcolor{red}{19.69}} & {\cellcolor[rgb]{0.792,0.918,1}}7.24\textsuperscript{\textcolor{red}{0.84}}  & {\cellcolor[rgb]{0.792,0.918,1}}13.90\textsuperscript{\textcolor{red}{0.22}}  & {\cellcolor[rgb]{0.792,0.918,1}}8.53\textsuperscript{\textcolor{red}{3.27}}   & {\cellcolor[rgb]{0.792,0.918,1}}20.10\textsuperscript{\textcolor{red}{5.51}}            & {\cellcolor[rgb]{0.792,0.918,1}}17.82\textsuperscript{\textcolor{red}{5.24}}  & {\cellcolor[rgb]{0.792,0.918,1}}40.17\textsuperscript{\textcolor{red}{6.37}}             \\
                                     & {\cellcolor[rgb]{0.792,0.918,1}}Oracle/E                 & {\cellcolor[rgb]{0.792,0.918,1}}24.04\textsuperscript{\textcolor{red}{14.76}} & {\cellcolor[rgb]{0.792,0.918,1}}44.71\textsuperscript{\textcolor{red}{17.08}} & {\cellcolor[rgb]{0.792,0.918,1}}6.80\textsuperscript{\textcolor{red}{0.40}}  & {\cellcolor[rgb]{0.792,0.918,1}}19.51\textsuperscript{\textcolor{red}{5.83}}  & {\cellcolor[rgb]{0.792,0.918,1}}6.24\textsuperscript{\textcolor{red}{0.98}}   & {\cellcolor[rgb]{0.792,0.918,1}}18.10\textsuperscript{\textcolor{red}{3.51}}            & {\cellcolor[rgb]{0.792,0.918,1}}13.38\textsuperscript{\textcolor{red}{0.80}}  & {\cellcolor[rgb]{0.792,0.918,1}}33.84\textsuperscript{\textcolor{red}{0.04}}             \\
                                     & {\cellcolor[rgb]{0.792,0.918,1}}Oracle/BE                & {\cellcolor[rgb]{0.792,0.918,1}}25.00\textsuperscript{\textcolor{red}{15.72}} & {\cellcolor[rgb]{0.792,0.918,1}}46.99\textsuperscript{\textcolor{red}{19.36}} & {\cellcolor[rgb]{0.792,0.918,1}}8.34\textsuperscript{\textcolor{red}{1.94}}  & {\cellcolor[rgb]{0.792,0.918,1}}20.41\textsuperscript{\textcolor{red}{6.73}}  & {\cellcolor[rgb]{0.792,0.918,1}}6.99\textsuperscript{\textcolor{red}{1.73}}   & {\cellcolor[rgb]{0.792,0.918,1}}16.45\textsuperscript{\textcolor{red}{1.86}}            & {\cellcolor[rgb]{0.792,0.918,1}}13.38\textsuperscript{\textcolor{red}{0.80}}  & {\cellcolor[rgb]{0.792,0.918,1}}29.21\textsuperscript{\textcolor[rgb]{0,0.502,0}{4.59}}  \\
                                     & Top1/B                                                   & 19.34\textsuperscript{\textcolor{red}{10.06}}                                 & 32.53\textsuperscript{\textcolor{red}{4.90}}                                  & 6.96\textsuperscript{\textcolor{red}{0.56}}                                  & 12.55\textsuperscript{\textcolor[rgb]{0,0.502,0}{1.13}}                       & 7.45\textsuperscript{\textcolor{red}{2.19}}                                   & 19.62\textsuperscript{\textcolor{red}{5.03}}                                            & 15.84\textsuperscript{\textcolor{red}{3.26}}                                  & 36.30\textsuperscript{\textcolor{red}{2.50}}                                             \\
                                     & Top2/B                                                   & 19.36\textsuperscript{\textcolor{red}{10.08}}                                 & 38.15\textsuperscript{\textcolor{red}{10.52}}                                 & 6.48\textsuperscript{\textcolor{red}{0.08}}                                  & 13.98\textsuperscript{\textcolor{red}{0.30}}                                  & 9.04\textsuperscript{\textcolor{red}{3.78}}                                   & 23.51\textsuperscript{\textcolor{red}{8.92}}                                            & 14.16\textsuperscript{\textcolor{red}{1.58}}                                  & 34.90\textsuperscript{\textcolor{red}{1.10}}                                             \\
                                     & Top2/E                                                   & 22.72\textsuperscript{\textcolor{red}{13.44}}                                 & 36.78\textsuperscript{\textcolor{red}{9.15}}                                  & 7.04\textsuperscript{\textcolor{red}{0.64}}                                  & 15.67\textsuperscript{\textcolor{red}{1.99}}                                  & 7.52\textsuperscript{\textcolor{red}{2.26}}                                   & 15.76\textsuperscript{\textcolor{red}{1.17}}                                            & 9.21\textsuperscript{\textcolor[rgb]{0,0.502,0}{3.37}}                        & 24.75\textsuperscript{\textcolor{red}{9.05}}                                             \\
                                     & Top2/BE                                                  & 21.60\textsuperscript{\textcolor{red}{12.32}}                                 & 34.04\textsuperscript{\textcolor{red}{6.41}}                                  & 7.90\textsuperscript{\textcolor{red}{1.50}}                                  & 20.40\textsuperscript{\textcolor{red}{6.72}}                                  & 7.04\textsuperscript{\textcolor{red}{1.78}}                                   & 16.94\textsuperscript{\textcolor{red}{2.35}}                                            & 8.19\textsuperscript{\textcolor[rgb]{0,0.502,0}{4.39}}                        & 19.54\textsuperscript{\textcolor{red}{14.26}}                                            \\
                                     & Top3/B                                                   & 22.10\textsuperscript{\textcolor{red}{12.82}}                                 & 36.81\textsuperscript{\textcolor{red}{9.18}}                                  & 5.76\textsuperscript{\textcolor[rgb]{0,0.502,0}{0.64}}                       & 11.45\textsuperscript{\textcolor[rgb]{0,0.502,0}{2.23}}                       & 8.79\textsuperscript{\textcolor{red}{3.53}}                                   & 22.93\textsuperscript{\textcolor{red}{8.34}}                                            & 13.55\textsuperscript{\textcolor{red}{0.97}}                                  & 34.64\textsuperscript{\textcolor{red}{0.84}}                                             \\
                                     & Top5/B                                                   & 18.20\textsuperscript{\textcolor{red}{8.92}}                                  & 32.27\textsuperscript{\textcolor{red}{4.64}}                                  & 6.49\textsuperscript{\textcolor{red}{0.09}}                                  & 13.96\textsuperscript{\textcolor{red}{0.28}}                                  & 10.56\textsuperscript{\textcolor{red}{5.3}}                                   & 24.16\textsuperscript{\textcolor{red}{9.57}}                                            & 14.09\textsuperscript{\textcolor{red}{1.51}}                                  & 35.11\textsuperscript{\textcolor{red}{1.31}}                                             \\
                                     & {\cellcolor[rgb]{1,0.941,0.835}}Human/B                  & {\cellcolor[rgb]{1,0.941,0.835}}19.32\textsuperscript{\textcolor{red}{10.04}} & {\cellcolor[rgb]{1,0.941,0.835}}36.40\textsuperscript{\textcolor{red}{8.77}}  & {\cellcolor[rgb]{1,0.941,0.835}}7.35\textsuperscript{\textcolor{red}{0.95}}  & {\cellcolor[rgb]{1,0.941,0.835}}14.63\textsuperscript{\textcolor{red}{0.95}}  & {\cellcolor[rgb]{1,0.941,0.835}}9.52\textsuperscript{\textcolor{red}{4.26}}   & {\cellcolor[rgb]{1,0.941,0.835}}21.52\textsuperscript{\textcolor{red}{6.93}}            & {\cellcolor[rgb]{1,0.941,0.835}}18.31\textsuperscript{\textcolor{red}{5.73}}  & {\cellcolor[rgb]{1,0.941,0.835}}39.44\textsuperscript{\textcolor{red}{5.64}}             \\ 
\hline
\multirow{11}{*}{Codex}           & No API                                                   & 8.08                                                                          & 24.47                                                                         & 6.12                                                                         & 12.58                                                                         & 3.40                                                                          & 10.84                                                                                   & 20.18                                                                         & 60.28                                                                                    \\
                                     & {\cellcolor[rgb]{0.792,0.918,1}}Oracle/B                 & {\cellcolor[rgb]{0.792,0.918,1}}44.10\textsuperscript{\textcolor{red}{36.02}} & {\cellcolor[rgb]{0.792,0.918,1}}73.07\textsuperscript{\textcolor{red}{48.60}} & {\cellcolor[rgb]{0.792,0.918,1}}9.02\textsuperscript{\textcolor{red}{2.90}}  & {\cellcolor[rgb]{0.792,0.918,1}}19.31\textsuperscript{\textcolor{red}{6.73}}  & {\cellcolor[rgb]{0.792,0.918,1}}15.93\textsuperscript{\textcolor{red}{12.53}} & {\cellcolor[rgb]{0.792,0.918,1}}45.60\textsuperscript{\textcolor{red}{34.76}}           & {\cellcolor[rgb]{0.792,0.918,1}}31.54\textsuperscript{\textcolor{red}{11.36}} & {\cellcolor[rgb]{0.792,0.918,1}}68.65\textsuperscript{\textcolor{red}{8.37}}             \\
                                     & {\cellcolor[rgb]{0.792,0.918,1}}Oracle/E                 & {\cellcolor[rgb]{0.792,0.918,1}}38.18\textsuperscript{\textcolor{red}{30.10}} & {\cellcolor[rgb]{0.792,0.918,1}}68.10\textsuperscript{\textcolor{red}{43.63}} & {\cellcolor[rgb]{0.792,0.918,1}}12.6\textsuperscript{\textcolor{red}{26.50}} & {\cellcolor[rgb]{0.792,0.918,1}}29.77\textsuperscript{\textcolor{red}{17.19}} & {\cellcolor[rgb]{0.792,0.918,1}}16.16\textsuperscript{\textcolor{red}{12.76}} & {\cellcolor[rgb]{0.792,0.918,1}}45.00\textsuperscript{\textcolor{red}{34.16}}           & {\cellcolor[rgb]{0.792,0.918,1}}27.80\textsuperscript{\textcolor{red}{7.62}}  & {\cellcolor[rgb]{0.792,0.918,1}}66.24\textsuperscript{\textcolor{red}{5.96}}             \\
                                     & {\cellcolor[rgb]{0.792,0.918,1}}Oracle/BE                & {\cellcolor[rgb]{0.792,0.918,1}}44.80\textsuperscript{\textcolor{red}{36.72}} & {\cellcolor[rgb]{0.792,0.918,1}}67.57\textsuperscript{\textcolor{red}{43.10}} & {\cellcolor[rgb]{0.792,0.918,1}}15.80\textsuperscript{\textcolor{red}{9.68}} & {\cellcolor[rgb]{0.792,0.918,1}}29.66\textsuperscript{\textcolor{red}{17.08}} & {\cellcolor[rgb]{0.792,0.918,1}}19.16\textsuperscript{\textcolor{red}{15.76}} & {\cellcolor[rgb]{0.792,0.918,1}}53.96\textsuperscript{\textcolor{red}{43.12}}           & {\cellcolor[rgb]{0.792,0.918,1}}32.84\textsuperscript{\textcolor{red}{12.66}} & {\cellcolor[rgb]{0.792,0.918,1}}71.50\textsuperscript{\textcolor{red}{11.22}}            \\
                                     & Top1/B                                                   & 14.82\textsuperscript{\textcolor{red}{6.74}}                                  & 38.07\textsuperscript{\textcolor{red}{13.60}}                                 & 7.34\textsuperscript{\textcolor{red}{1.22}}                                  & 16.42\textsuperscript{\textcolor{red}{3.84}}                                  & 13.73\textsuperscript{\textcolor{red}{10.33}}                                 & 39.22\textsuperscript{\textcolor{red}{28.38}}                                           & 24.88\textsuperscript{\textcolor{red}{4.70}}                                  & 58.66\textsuperscript{\textcolor[rgb]{0,0.502,0}{1.62}}                                  \\
                                     & Top2/B                                                   & 19.06\textsuperscript{\textcolor{red}{10.98}}                                 & 49.70\textsuperscript{\textcolor{red}{25.23}}                                 & 6.78\textsuperscript{\textcolor{red}{0.66}}                                  & 15.60\textsuperscript{\textcolor{red}{3.02}}                                  & 15.96\textsuperscript{\textcolor{red}{12.56}}                                 & 41.76\textsuperscript{\textcolor{red}{30.92}}                                           & 27.09\textsuperscript{\textcolor{red}{6.91}}                                  & 60.76\textsuperscript{\textcolor{red}{0.48}}                                             \\
                                     & Top2/E                                                   & 23.94\textsuperscript{\textcolor{red}{15.86}}                                 & 46.26\textsuperscript{\textcolor{red}{21.79}}                                 & 11.12\textsuperscript{\textcolor{red}{5.00}}                                 & 21.17\textsuperscript{\textcolor{red}{8.59}}                                  & 14.98\textsuperscript{\textcolor{red}{11.58}}                                 & 41.41\textsuperscript{\textcolor{red}{30.57}}                                           & 20.97\textsuperscript{\textcolor{red}{0.79}}                                  & 57.75\textsuperscript{\textcolor[rgb]{0,0.502,0}{2.53}}                                  \\
                                     & Top2/BE                                                  & 24.82\textsuperscript{\textcolor{red}{16.74}}                                 & 51.43\textsuperscript{\textcolor{red}{26.96}}                                 & 13.32\textsuperscript{\textcolor{red}{7.20}}                                 & 21.64\textsuperscript{\textcolor{red}{9.06}}                                  & 15.69\textsuperscript{\textcolor{red}{12.29}}                                 & 42.40\textsuperscript{\textcolor{red}{31.56}}                                           & 20.26\textsuperscript{\textcolor{red}{0.08}}                                  & 54.96\textsuperscript{\textcolor[rgb]{0,0.502,0}{5.32}}                                  \\
                                     & Top3/B                                                   & 19.32\textsuperscript{\textcolor{red}{11.24}}                                 & 49.55\textsuperscript{\textcolor{red}{25.08}}                                 & 7.62\textsuperscript{\textcolor{red}{1.50}}                                  & 18.75\textsuperscript{\textcolor{red}{6.17}}                                  & 13.27\textsuperscript{\textcolor{red}{9.87}}                                  & 41.19\textsuperscript{\textcolor{red}{30.35}}                                           & 26.96\textsuperscript{\textcolor{red}{6.78}}                                  & 62.63\textsuperscript{\textcolor{red}{2.35}}                                             \\
                                     & Top5/B                                                   & 20.36\textsuperscript{\textcolor{red}{12.28}}                                 & 52.71\textsuperscript{\textcolor{red}{28.24}}                                 & 7.18\textsuperscript{\textcolor{red}{1.06}}                                  & 16.33\textsuperscript{\textcolor{red}{3.75}}                                  & 17.00\textsuperscript{\textcolor{red}{13.60}}                                 & 45.94\textsuperscript{\textcolor{red}{35.10}}                                           & 26.81\textsuperscript{\textcolor{red}{6.63}}                                  & 63.52\textsuperscript{\textcolor{red}{3.24}}                                             \\
                                     & {\cellcolor[rgb]{1,0.941,0.835}}Human/B                  & {\cellcolor[rgb]{1,0.941,0.835}}15.24\textsuperscript{\textcolor{red}{7.16}}  & {\cellcolor[rgb]{1,0.941,0.835}}40.68\textsuperscript{\textcolor{red}{16.21}} & {\cellcolor[rgb]{1,0.941,0.835}}6.98\textsuperscript{\textcolor{red}{0.86}}  & {\cellcolor[rgb]{1,0.941,0.835}}18.57\textsuperscript{\textcolor{red}{5.99}}  & {\cellcolor[rgb]{1,0.941,0.835}}14.97\textsuperscript{\textcolor{red}{11.57}} & {\cellcolor[rgb]{1,0.941,0.835}}39.88\textsuperscript{\textcolor{red}{29.04}}           & {\cellcolor[rgb]{1,0.941,0.835}}29.19\textsuperscript{\textcolor{red}{9.01}}  & {\cellcolor[rgb]{1,0.941,0.835}}64.25\textsuperscript{\textcolor{red}{3.97}}             \\
\bottomrule
\end{tabular}
}
\end{table*}

\section{Experiments}
In this section, we will showcase a thorough evaluation of our proposed methodology through a sequence of carefully designed experiments. Specifically, we will initially outline the experimental setup, subsequently delving into a comprehensive presentation of our findings.

\subsection{Experimental Setup}

\subsubsection{Baselines}
Our contributions can be viewed from two perspectives: \apifinder and \apicoder.
With regards to \apifinder, we propose a retrieval-augmented generative model for code generation in private libraries.
Therefore, the baselines are all code generation models with No API setting, while we propose Oracle, Top$K$, and Human settings.
From the perspective of \apicoder, we propose a novel idea to build an advanced model by continually pre-training the vanilla one.
So, vanilla models are the baselines for their advanced versions.
Overall, we do comprehensive comparisons among $17$ popular code generation models, such as \codegen~\cite{codegen}, GPT-CC~\cite{gptcc}, InCoder~\cite{incoder}, CodeGPT~\cite{codegpt}, CodeT5~\cite{codet5}, CodeParrot~\cite{codeparrot}, SantaCoder~\cite{santacoder}, PyCodeGPT~\cite{CERT}, PolyCoder~\cite{polycoder}, and OpenAI's code-davinci-002~\cite{codex}.
For the sake of brevity, we abbreviate the code-davinci-002 model as Codex in the following sections.

\subsubsection{Evaluation Metrics}
We adopt pass$@k$ as our evaluation metric in accordance with Codex~\cite{codex}.
For each problem, we sample $n=100$ candidate code snippets from the LLM. Then, we count the number $c$ of correct ones by running on test cases.
Pass$@k$ can be formalized as:
\begin{equation}
\passk =
\begin{cases}
1 &\text{if } n - c < k \\
1 - \prod_{i=n-c+1}^{n} (1 - \frac{k}{i}) &\text{otherwise}
\end{cases},
\end{equation}
where $k \in \{1,10,100\}$ in our study.
Besides pass$@k$, other metrics also exist, such as ROUGE~\cite{rouge}, BLEU~\cite{bleu}, and CodeBLEU~\cite{codebleu}.
We chose pass$@k$ as our evaluation metric instead of others because it can provide a completely precise evaluation of code accuracy by executing test cases, while others do not.

\subsubsection{Implementation Details}
We use Dense\footnote{\url{https://github.com/luyug/Dense}}, a toolkit for training dense retriever, to implement the \apifinder.
Regarding the training details of \apifinder, the ratio of positive to negative samples is set to $1$:$8$, with a batch size of $10$ per GPU card, a learning rate of $1$e-$5$ and optimization via the Adam algorithm~\cite{adam}.
Our training duration was approximately $74$ hours on an $8$ GPU NVIDIA V$100$ ($32$GB) cluster, with a total of $100$K steps.
For \codegenapi, we totally train $9$ versions with \codegen $350$M, $2$B, and $6$B, each training the three prompts depicted in Figure~\ref{fig:apicoder}.
These versions exactly follow the original hyperparameters of \codegen for continuous pre-training.
We use DeepSpeed~\cite{deepspeed} to train \codegenapi with FP$16$ precision and employ an $32$ GPU NVIDIA A100 ($48$GB) cluster.
During the inference of all \apicoder models, the number of samples is set to $100$, the temperature to $0.8$, the maximum length of new sequences to $300$, and the top-p to $0.95$.

\subsubsection{Experimental Configurations}
\textit{Which APIs should be input to \apicoder for a programming problem?}
In our experiment, we provide the following four settings:
\begin{itemize}
    \item No API: no API is input to \apicoder.
    \item Oracle: the ground-truth APIs are input to \apicoder.
    \item Top$K$: The first $K$ APIs retrieved by \apifinder are input to \apicoder, where $K\in\{1,2,3,5\}$.
    \item Human: \apifinder offers the top $5$ APIs for user selection (Figure~\ref{fig:human_in_the_loop}), and the APIs chosen by a user are input to \apicoder.
\end{itemize}

\textit{Which components in the \apidoc should be placed in the prompt?}
In fact, we find that some components in \apidoc, including API parameters and related APIs, significantly devastate the performance of LLMs.
We thus only consider the following three types of prompts:
\begin{itemize}
    \item API Basic Only (B): the prompt only includes API basic consisting of API name, signature, and description.
    \item API Examples Only (E): the prompt only includes API examples.
    \item API Basic and Examples (BE): the prompt includes both API basic and examples.
\end{itemize}

\begin{table}[width=1.0\linewidth,pos=t]
\centering
\caption{\Passk (\%) of various components of \apidoc using \codegen $350$M in the ``Oracle'' setting.}
\label{tab:api_doc_components_results}
\resizebox{1.0\linewidth}{!}{
\begin{tabular}{lllll} 
\toprule
\multirow{2}{*}{\textbf{API Doc.}} & \multicolumn{2}{c}{\textbf{TorchDataEval}} & \multicolumn{2}{c}{\begin{tabular}[c]{@{}c@{}}\textbf{TorchData}\\\textbf{ComplexEval}\end{tabular}}  \\ 
\cline{2-5}
                                   & pass@1 & pass@10                           & pass@1 & pass@10                                                                                      \\ 
\hline\hline
Basic                              & 7.22   & 28.53                             & 1.36   & 6.52                                                                                         \\
Examples                           & 13.12  & 30.84                             & 1.92   & 6.71                                                                                         \\
Parameters                         & 3.95   & 14.13                             & 0.85   & 3.22                                                                                         \\
Related APIs                       & 4.15   & 14.48                             & 0.86   & 4.46                                                                                         \\
\bottomrule
\end{tabular}
}
\end{table}

\subsection{Main Results} \label{sec:main_results}
In this section, we focus on four essential research questions (RQs) to evaluate the effectiveness of our proposed approach in the private library scenario.

\paragraph{RQ1: ``Whether off-the-shelf LLMs possess the potential to invoke private APIs?''}
This aims to verify the feasibility of LLMs in addressing the private library scenario.
Specifically, we furnish these LLMs with the ground truth (oracle) private APIs and assess their capability to invoke them correctly.
Table~\ref{tab:main_results} reveals a substantial improvement when providing oracle APIs compared to no APIs, indicating the potential of LLMs for invoking private APIs.
Intriguingly, even the giant model, codex, performs poorly in invoking private APIs without prompting any APIs, further emphasizing the necessity of our approach.
Excitingly, as the parameter size increase (\codegen $350$M < \codegen $2$B < \codegen $6$B < Codex), the benefits from prompting APIs grow commensurately.
For instance, Codex yields a $48.60$\% pass@10 gain in \torchdataeval in the ``Oracle/B'' setting, while \codegen $350$M merely achieves a $10.77$\% increase.

\paragraph{RQ2: ``Which components in API documentation are more useful for LLMs?''}
As previously shown, LLMs possess the potential to invoke the private APIs via prompt API information in \apidoc. So, which information in \apidoc is crucial for maximizing the performance of LLMs? We separately prompt each component mentioned in Section~\ref{sec:api_documentation} to LLM. The results in Table~\ref{tab:api_doc_components_results}  demonstrate that API basic and examples are more beneficial compared to API parameters and related APIs. This is reasonable as the former two components directly provide definitions or invocations for the API, while the latter two do not. Therefore, in this paper, we primarily focus on API basic and examples.

\begin{table*}[width=2.0\linewidth,pos=h]
\centering
\caption{\Passk (\%) of our proposed \codegenapi on four private library benchmarks. \#1, \#2, and \#3 denote \codegenapi trained with the three prompts in Figure~\ref{fig:apicoder}. \textcolor{red}{Red} and \textcolor{darkgreen}{green} values indicate improvements and deteriorations of \codegenapi compared to \codegen under the same settings, respectively.}
\label{tab:analysis_results}
\resizebox{\linewidth}{!}{
\begin{tabular}{c:l:l|ll:ll:ll:ll} 
\toprule
\multicolumn{1}{l}{\multirow{2}{*}{\textbf{APICoder }}}                 & \multicolumn{1}{l}{\multirow{2}{*}{\textbf{P.}}} & \multicolumn{1}{l}{\multirow{2}{*}{\textbf{APIFinder }}} & \multicolumn{2}{c}{\textbf{TorchDataEval }}                                                                                                                   & \multicolumn{2}{c}{\begin{tabular}[c]{@{}c@{}}\textbf{TorchData}\\\textbf{ComplexEval }\end{tabular}}                                                       & \multicolumn{2}{c}{\textbf{ MonkeyEval}}                                                                                                                      & \multicolumn{2}{c}{\textbf{ BeatNumEval}}                                                                                                                                \\ 
\cline{4-11}
\multicolumn{1}{l}{}                                                    & \multicolumn{1}{l}{}                             & \multicolumn{1}{l}{}                                     & pass@1                                                                        & \multicolumn{1}{l}{pass@10}                                                   & pass@1                                                                       & \multicolumn{1}{l}{pass@10}                                                  & pass@1                                                                        & \multicolumn{1}{l}{pass@10}                                                   & pass@1                                                                                  & pass@10                                                                        \\ 
\hline\hline
\multirow{10}{*}{\begin{tabular}[c]{@{}c@{}}\codegenapi\\350M\end{tabular}} & \multirow{6}{*}{\#1}                             & Oracle/B                                                 & {\cellcolor[rgb]{0.792,0.918,1}}16.54\textsuperscript{\textcolor{red}{9.32}}  & {\cellcolor[rgb]{0.792,0.918,1}}38.95\textsuperscript{\textcolor{red}{10.42}} & {\cellcolor[rgb]{0.792,0.918,1}}2.16\textsuperscript{\textcolor{red}{0.80}}  & {\cellcolor[rgb]{0.792,0.918,1}}8.10\textsuperscript{\textcolor{red}{1.58}}  & {\cellcolor[rgb]{0.792,0.918,1}}5.81\textsuperscript{\textcolor{red}{2.52}}   & {\cellcolor[rgb]{0.792,0.918,1}}15.27\textsuperscript{\textcolor{red}{3.89}}  & {\cellcolor[rgb]{0.792,0.918,1}}12.53\textsuperscript{\textcolor{red}{1.34}}            & {\cellcolor[rgb]{0.792,0.918,1}}31.97\textsuperscript{\textcolor{red}{3.05}}   \\
                                                                        &                                                  & Top1/B                                                   & 11.53\textsuperscript{\textcolor{red}{5.49}}                                  & 27.76\textsuperscript{\textcolor{red}{6.03}}                                  & 2.02\textsuperscript{\textcolor{red}{0.74}}                                  & 7.57\textsuperscript{\textcolor{red}{1.36}}                                  & 6.25\textsuperscript{\textcolor{red}{2.88}}                                   & 15.05\textsuperscript{\textcolor{red}{4.03}}                                  & 10.33\textsuperscript{\textcolor{red}{1.46}}                                            & 27.63\textsuperscript{\textcolor{red}{3.35}}                                   \\
                                                                        &                                                  & Top2/B                                                   & 9.09\textsuperscript{\textcolor{red}{3.37}}                                   & 24.92\textsuperscript{\textcolor{red}{4.56}}                                  & 2.18\textsuperscript{\textcolor{red}{0.82}}                                  & 8.42\textsuperscript{\textcolor{red}{1.22}}                                  & 6.21\textsuperscript{\textcolor{red}{2.45}}                                   & 17.01\textsuperscript{\textcolor{red}{4.62}}                                  & 10.35\textsuperscript{\textcolor{red}{1.04}}                                            & 26.89\textsuperscript{\textcolor{red}{2.56}}                                   \\
                                                                        &                                                  & Top3/B                                                   & 9.49\textsuperscript{\textcolor{red}{3.21}}                                   & 25.31\textsuperscript{\textcolor{red}{3.25}}                                  & 1.99\textsuperscript{\textcolor{red}{0.79}}                                  & 8.61\textsuperscript{\textcolor{red}{1.12}}                                  & 6.20\textsuperscript{\textcolor{red}{2.36}}                                   & 15.07\textsuperscript{\textcolor{red}{3.46}}                                  & 10.32\textsuperscript{\textcolor{red}{1.25}}                                            & 28.95\textsuperscript{\textcolor{red}{3.13}}                                   \\
                                                                        &                                                  & Top5/B                                                   & 11.03\textsuperscript{\textcolor{red}{3.99}}                                  & 28.95\textsuperscript{\textcolor{red}{3.94}}                                  & 1.54\textsuperscript{\textcolor{red}{0.42}}                                  & 7.46\textsuperscript{\textcolor{red}{0.83}}                                  & 5.58\textsuperscript{\textcolor{red}{1.50}}                                   & 15.48\textsuperscript{\textcolor{red}{2.74}}                                  & 9.96\textsuperscript{\textcolor{red}{1.03}}                                             & 29.30\textsuperscript{\textcolor{red}{3.04}}                                   \\
                                                                        &                                                  & Human/B                                                  & 14.97\textsuperscript{\textcolor{red}{8.35}}                                  & 35.86\textsuperscript{\textcolor{red}{9.83}}                                  & 2.58\textsuperscript{\textcolor{red}{1.14}}                                  & 8.75\textsuperscript{\textcolor{red}{1.94}}                                  & 6.30\textsuperscript{\textcolor{red}{2.95}}                                   & 16.45\textsuperscript{\textcolor{red}{4.67}}                                  & 12.92\textsuperscript{\textcolor{red}{1.53}}                                            & 33.56\textsuperscript{\textcolor{red}{3.44}}                                   \\ 
\cdashline{2-2}
                                                                        & \multirow{2}{*}{\#2}                             & Oracle/E                                                 & {\cellcolor[rgb]{0.792,0.918,1}}26.64\textsuperscript{\textcolor{red}{13.52}} & {\cellcolor[rgb]{0.792,0.918,1}}47.16\textsuperscript{\textcolor{red}{16.32}} & {\cellcolor[rgb]{0.792,0.918,1}}3.17\textsuperscript{\textcolor{red}{1.25}}  & {\cellcolor[rgb]{0.792,0.918,1}}8.66\textsuperscript{\textcolor{red}{1.95}}  & {\cellcolor[rgb]{0.792,0.918,1}}6.55\textsuperscript{\textcolor{red}{3.14}}   & {\cellcolor[rgb]{0.792,0.918,1}}20.56\textsuperscript{\textcolor{red}{5.15}}  & {\cellcolor[rgb]{0.792,0.918,1}}6.58\textsuperscript{\textcolor{red}{0.05}}             & {\cellcolor[rgb]{0.792,0.918,1}}24.14\textsuperscript{\textcolor{red}{0.96}}   \\
                                                                        &                                                  & Top2/E                                                   & 11.80\textsuperscript{\textcolor{red}{5.14}}                                  & 23.65\textsuperscript{\textcolor{red}{7.26}}                                  & 2.11\textsuperscript{\textcolor{red}{0.99}}                                  & 8.27\textsuperscript{\textcolor{red}{1.52}}                                  & 4.97\textsuperscript{\textcolor{red}{2.55}}                                   & 15.53\textsuperscript{\textcolor{red}{4.69}}                                  & 5.43\textsuperscript{\textcolor[rgb]{0,0.502,0}{-0.07}}                                 & 16.62\textsuperscript{\textcolor{red}{0.25}}                                   \\ 
\cdashline{2-2}
                                                                        & \multirow{2}{*}{\#3}                             & Oracle/BE                                                & {\cellcolor[rgb]{0.792,0.918,1}}23.51\textsuperscript{\textcolor{red}{11.45}} & {\cellcolor[rgb]{0.792,0.918,1}}45.56\textsuperscript{\textcolor{red}{13.90}} & {\cellcolor[rgb]{0.792,0.918,1}}3.00\textsuperscript{\textcolor{red}{1.24}}  & {\cellcolor[rgb]{0.792,0.918,1}}10.67\textsuperscript{\textcolor{red}{2.05}} & {\cellcolor[rgb]{0.792,0.918,1}}5.55\textsuperscript{\textcolor{red}{2.90}}   & {\cellcolor[rgb]{0.792,0.918,1}}16.46\textsuperscript{\textcolor{red}{4.85}}  & {\cellcolor[rgb]{0.792,0.918,1}}7.87\textsuperscript{\textcolor{red}{0.35}}             & {\cellcolor[rgb]{0.792,0.918,1}}26.70\textsuperscript{\textcolor{red}{1.89}}   \\
                                                                        &                                                  & Top2/BE                                                  & 10.28\textsuperscript{\textcolor{red}{4.84}}                                  & 23.26\textsuperscript{\textcolor{red}{5.16}}                                  & 2.50\textsuperscript{\textcolor{red}{1.06}}                                  & 9.62\textsuperscript{\textcolor{red}{1.96}}                                  & 3.03\textsuperscript{\textcolor{red}{1.04}}                                   & 12.33\textsuperscript{\textcolor{red}{3.11}}                                  & 4.42\textsuperscript{\textcolor{red}{0.14}}                                             & 16.31\textsuperscript{\textcolor{red}{1.58}}                                   \\ 
\hline
\multirow{10}{*}{\begin{tabular}[c]{@{}c@{}}\codegenapi\\2B\end{tabular}}   & \multirow{6}{*}{\#1}                             & Oracle/B                                                 & {\cellcolor[rgb]{0.792,0.918,1}}33.32\textsuperscript{\textcolor{red}{14.52}} & {\cellcolor[rgb]{0.792,0.918,1}}59.18\textsuperscript{\textcolor{red}{16.25}} & {\cellcolor[rgb]{0.792,0.918,1}}8.34\textsuperscript{\textcolor{red}{2.74}}  & {\cellcolor[rgb]{0.792,0.918,1}}17.84\textsuperscript{\textcolor{red}{4.25}} & {\cellcolor[rgb]{0.792,0.918,1}}13.52\textsuperscript{\textcolor{red}{6.29}}  & {\cellcolor[rgb]{0.792,0.918,1}}25.60\textsuperscript{\textcolor{red}{8.35}}  & {\cellcolor[rgb]{0.792,0.918,1}}20.47\textsuperscript{\textcolor{red}{3.74}}            & {\cellcolor[rgb]{0.792,0.918,1}}45.14\textsuperscript{\textcolor{red}{7.35}}   \\
                                                                        &                                                  & Top1/B                                                   & 20.30\textsuperscript{\textcolor{red}{10.10}}                                 & 39.53\textsuperscript{\textcolor{red}{11.55}}                                 & 8.26\textsuperscript{\textcolor{red}{2.66}}                                  & 17.65\textsuperscript{\textcolor{red}{4.04}}                                 & 14.67\textsuperscript{\textcolor{red}{7.74}}                                  & 27.29\textsuperscript{\textcolor{red}{9.42}}                                  & 16.60\textsuperscript{\textcolor{red}{3.53}}                                            & 38.46\textsuperscript{\textcolor{red}{6.88}}                                   \\
                                                                        &                                                  & Top2/B                                                   & 20.45\textsuperscript{\textcolor{red}{7.45}}                                  & 44.14\textsuperscript{\textcolor{red}{9.10}}                                  & 8.60\textsuperscript{\textcolor{red}{2.80}}                                  & 17.57\textsuperscript{\textcolor{red}{3.79}}                                 & 13.67\textsuperscript{\textcolor{red}{5.25}}                                  & 28.34\textsuperscript{\textcolor{red}{8.22}}                                  & 15.94\textsuperscript{\textcolor{red}{2.97}}                                            & 36.85\textsuperscript{\textcolor{red}{6.03}}                                   \\
                                                                        &                                                  & Top3/B                                                   & 19.79\textsuperscript{\textcolor{red}{7.19}}                                  & 45.61\textsuperscript{\textcolor{red}{9.42}}                                  & 7.16\textsuperscript{\textcolor{red}{2.76}}                                  & 16.24\textsuperscript{\textcolor{red}{3.63}}                                 & 14.89\textsuperscript{\textcolor{red}{6.77}}                                  & 27.68\textsuperscript{\textcolor{red}{9.10}}                                  & 14.63\textsuperscript{\textcolor{red}{3.05}}                                            & 36.78\textsuperscript{\textcolor{red}{6.49}}                                   \\
                                                                        &                                                  & Top5/B                                                   & 18.75\textsuperscript{\textcolor{red}{8.15}}                                  & 44.52\textsuperscript{\textcolor{red}{10.57}}                                 & 7.14\textsuperscript{\textcolor{red}{1.94}}                                  & 16.73\textsuperscript{\textcolor{red}{3.13}}                                 & 14.92\textsuperscript{\textcolor{red}{5.51}}                                  & 27.65\textsuperscript{\textcolor{red}{7.42}}                                  & 13.74\textsuperscript{\textcolor{red}{1.96}}                                            & 37.12\textsuperscript{\textcolor{red}{4.96}}                                   \\
                                                                        &                                                  & Human/B                                                  & 24.37\textsuperscript{\textcolor{red}{12.17}}                                 & 45.07\textsuperscript{\textcolor{red}{15.56}}                                 & 8.73\textsuperscript{\textcolor{red}{2.85}}                                  & 18.22\textsuperscript{\textcolor{red}{4.62}}                                 & 14.17\textsuperscript{\textcolor{red}{7.04}}                                  & 25.71\textsuperscript{\textcolor{red}{8.70}}                                  & 20.00\textsuperscript{\textcolor{red}{3.76}}                                            & 42.05\textsuperscript{\textcolor{red}{6.28}}                                   \\ 
\cdashline{2-2}
                                                                        & \multirow{2}{*}{\#2}                             & Oracle/E                                                 & {\cellcolor[rgb]{0.792,0.918,1}}39.54\textsuperscript{\textcolor{red}{16.14}} & {\cellcolor[rgb]{0.792,0.918,1}}61.05\textsuperscript{\textcolor{red}{19.09}} & {\cellcolor[rgb]{0.792,0.918,1}}11.02\textsuperscript{\textcolor{red}{4.02}} & {\cellcolor[rgb]{0.792,0.918,1}}23.48\textsuperscript{\textcolor{red}{4.99}} & {\cellcolor[rgb]{0.792,0.918,1}}13.36\textsuperscript{\textcolor{red}{7.62}}  & {\cellcolor[rgb]{0.792,0.918,1}}23.29\textsuperscript{\textcolor{red}{9.21}}  & {\cellcolor[rgb]{0.792,0.918,1}}12.14\textsuperscript{\textcolor{red}{1.45}}            & {\cellcolor[rgb]{0.792,0.918,1}}27.50\textsuperscript{\textcolor{red}{1.04}}   \\
                                                                        &                                                  & Top2/E                                                   & 26.24\textsuperscript{\textcolor{red}{9.24}}                                  & 43.19\textsuperscript{\textcolor{red}{12.45}}                                 & 9.71\textcolor{red}{\textsuperscript{3.91}}                                  & 17.26\textsuperscript{\textcolor{red}{4.03}}                                 & 11.19\textsuperscript{\textcolor{red}{6.64}}                                  & 20.49\textsuperscript{\textcolor{red}{7.57}}                                  & 7.37\textsuperscript{\textcolor[rgb]{0,0.502,0}{-0.15}}                                 & 17.96\textsuperscript{\textcolor[rgb]{0,0.502,0}{-0.07}}                       \\ 
\cdashline{2-2}
                                                                        & \multirow{2}{*}{\#3}                             & Oracle/BE                                                & {\cellcolor[rgb]{0.792,0.918,1}}38.36\textsuperscript{\textcolor{red}{14.36}} & {\cellcolor[rgb]{0.792,0.918,1}}59.82\textsuperscript{\textcolor{red}{16.27}} & {\cellcolor[rgb]{0.792,0.918,1}}10.74\textsuperscript{\textcolor{red}{4.34}} & {\cellcolor[rgb]{0.792,0.918,1}}22.86\textsuperscript{\textcolor{red}{4.27}} & {\cellcolor[rgb]{0.792,0.918,1}}12.59\textsuperscript{\textcolor{red}{6.85}}  & {\cellcolor[rgb]{0.792,0.918,1}}17.62\textsuperscript{\textcolor{red}{6.13}}  & {\cellcolor[rgb]{0.792,0.918,1}}9.45\textsuperscript{\textcolor[rgb]{0,0.502,0}{-0.05}} & {\cellcolor[rgb]{0.792,0.918,1}}33.23\textsuperscript{\textcolor{red}{2.41}}   \\
                                                                        &                                                  & Top2/BE                                                  & 23.15\textsuperscript{\textcolor{red}{7.15}}                                  & 42.29\textsuperscript{\textcolor{red}{11.12}}                                 & 10.26\textsuperscript{\textcolor{red}{3.46}}                                 & 20.16\textsuperscript{\textcolor{red}{2.95}}                                 & 10.18\textsuperscript{\textcolor{red}{5.23}}                                  & 17.03\textsuperscript{\textcolor{red}{4.54}}                                  & 6.41\textsuperscript{\textcolor{red}{2.45}}                                             & 17.12\textsuperscript{\textcolor{red}{6.90}}                                   \\ 
\hline
\multirow{10}{*}{\begin{tabular}[c]{@{}c@{}}\codegenapi\\6B\end{tabular}}   & \multirow{6}{*}{\#1}                             & Oracle/B                                                 & {\cellcolor[rgb]{0.792,0.918,1}}45.23\textsuperscript{\textcolor{red}{20.51}} & {\cellcolor[rgb]{0.792,0.918,1}}71.41\textsuperscript{\textcolor{red}{24.09}} & {\cellcolor[rgb]{0.792,0.918,1}}13.26\textsuperscript{\textcolor{red}{6.02}} & {\cellcolor[rgb]{0.792,0.918,1}}23.54\textsuperscript{\textcolor{red}{9.64}} & {\cellcolor[rgb]{0.792,0.918,1}}17.57\textsuperscript{\textcolor{red}{9.04}}  & {\cellcolor[rgb]{0.792,0.918,1}}32.67\textsuperscript{\textcolor{red}{12.57}} & {\cellcolor[rgb]{0.792,0.918,1}}25.16\textsuperscript{\textcolor{red}{7.34}}            & {\cellcolor[rgb]{0.792,0.918,1}}53.04\textsuperscript{\textcolor{red}{12.87}}  \\
                                                                        &                                                  & Top1/B                                                   & 34.69\textsuperscript{\textcolor{red}{15.35}}                                 & 49.67\textsuperscript{\textcolor{red}{17.14}}                                 & 13.42\textsuperscript{\textcolor{red}{6.46}}                                 & 21.85\textsuperscript{\textcolor{red}{9.30}}                                 & 18.97\textsuperscript{\textcolor{red}{11.52}}                                 & 34.48\textsuperscript{\textcolor{red}{14.86}}                                 & 21.93\textsuperscript{\textcolor{red}{6.09}}                                            & 45.75\textsuperscript{\textcolor{red}{9.45}}                                   \\
                                                                        &                                                  & Top2/B                                                   & 33.98\textsuperscript{\textcolor{red}{14.62}}                                 & 53.92\textsuperscript{\textcolor{red}{15.77}}                                 & 11.74\textsuperscript{\textcolor{red}{5.26}}                                 & 22.23\textsuperscript{\textcolor{red}{8.25}}                                 & 18.50\textsuperscript{\textcolor{red}{9.46}}                                  & 37.44\textsuperscript{\textcolor{red}{13.93}}                                 & 20.02\textsuperscript{\textcolor{red}{5.86}}                                            & 44.49\textsuperscript{\textcolor{red}{9.59}}                                   \\
                                                                        &                                                  & Top3/B                                                   & 35.35\textsuperscript{\textcolor{red}{13.25}}                                 & 51.34\textsuperscript{\textcolor{red}{14.53}}                                 & 12.01\textsuperscript{\textcolor{red}{6.25}}                                 & 18.78\textsuperscript{\textcolor{red}{7.33}}                                 & 19.36\textsuperscript{\textcolor{red}{10.57}}                                 & 35.74\textsuperscript{\textcolor{red}{12.81}}                                 & 19.57\textsuperscript{\textcolor{red}{6.02}}                                            & 43.15\textsuperscript{\textcolor{red}{8.51}}                                   \\
                                                                        &                                                  & Top5/B                                                   & 32.47\textsuperscript{\textcolor{red}{14.27}}                                 & 47.02\textsuperscript{\textcolor{red}{14.75}}                                 & 12.75\textsuperscript{\textcolor{red}{6.26}}                                 & 19.73\textsuperscript{\textcolor{red}{5.77}}                                 & 20.07\textsuperscript{\textcolor{red}{9.51}}                                  & 35.76\textsuperscript{\textcolor{red}{11.60}}                                 & 19.21\textsuperscript{\textcolor{red}{5.12}}                                            & 42.67\textsuperscript{\textcolor{red}{7.56}}                                   \\
                                                                        &                                                  & Human/B                                                  & 35.66\textsuperscript{\textcolor{red}{16.34}}                                 & 52.03\textsuperscript{\textcolor{red}{15.63}}                                 & 14.29\textsuperscript{\textcolor{red}{6.94}}                                 & 23.36\textsuperscript{\textcolor{red}{8.73}}                                 & 20.91\textsuperscript{\textcolor{red}{11.39}}                                 & 35.24\textsuperscript{\textcolor{red}{13.72}}                                 & 25.31\textsuperscript{\textcolor{red}{7.00}}                                            & 49.90\textsuperscript{\textcolor{red}{10.46}}                                  \\ 
\cdashline{2-2}
                                                                        & \multirow{2}{*}{\#2}                             & Oracle/E                                                 & {\cellcolor[rgb]{0.792,0.918,1}}46.60\textsuperscript{\textcolor{red}{22.56}} & {\cellcolor[rgb]{0.792,0.918,1}}69.85\textsuperscript{\textcolor{red}{25.14}} & {\cellcolor[rgb]{0.792,0.918,1}}15.77\textsuperscript{\textcolor{red}{8.97}} & {\cellcolor[rgb]{0.792,0.918,1}}26.62\textsuperscript{\textcolor{red}{7.11}} & {\cellcolor[rgb]{0.792,0.918,1}}16.27\textsuperscript{\textcolor{red}{10.03}} & {\cellcolor[rgb]{0.792,0.918,1}}29.56\textsuperscript{\textcolor{red}{11.46}} & {\cellcolor[rgb]{0.792,0.918,1}}19.14\textsuperscript{\textcolor{red}{5.76}}            & {\cellcolor[rgb]{0.792,0.918,1}}38.09\textsuperscript{\textcolor{red}{4.25}}   \\
                                                                        &                                                  & Top2/E                                                   & 37.86\textsuperscript{\textcolor{red}{15.14}}                                 & 53.45\textsuperscript{\textcolor{red}{16.67}}                                 & 14.50\textsuperscript{\textcolor{red}{7.46}}                                 & 23.62\textsuperscript{\textcolor{red}{7.95}}                                 & 16.98\textsuperscript{\textcolor{red}{9.46}}                                  & 25.75\textsuperscript{\textcolor{red}{9.99}}                                  & 9.78\textsuperscript{\textcolor{red}{0.57}}                                             & 26.21\textsuperscript{\textcolor{red}{1.46}}                                   \\ 
\cdashline{2-2}
                                                                        & \multirow{2}{*}{\#3}                             & Oracle/BE                                                & {\cellcolor[rgb]{0.792,0.918,1}}45.46\textsuperscript{\textcolor{red}{20.46}} & {\cellcolor[rgb]{0.792,0.918,1}}66.54\textsuperscript{\textcolor{red}{19.55}} & {\cellcolor[rgb]{0.792,0.918,1}}16.76\textsuperscript{\textcolor{red}{8.42}} & {\cellcolor[rgb]{0.792,0.918,1}}28.41\textsuperscript{\textcolor{red}{8.00}} & {\cellcolor[rgb]{0.792,0.918,1}}17.56\textsuperscript{\textcolor{red}{10.57}} & {\cellcolor[rgb]{0.792,0.918,1}}29.56\textsuperscript{\textcolor{red}{13.11}} & {\cellcolor[rgb]{0.792,0.918,1}}13.38\textsuperscript{\textcolor{red}{0.00}}            & {\cellcolor[rgb]{0.792,0.918,1}}30.73\textsuperscript{\textcolor{red}{1.52}}   \\
                                                                        &                                                  & Top2/BE                                                  & 36.85\textsuperscript{\textcolor{red}{15.25}}                                 & 52.37\textsuperscript{\textcolor{red}{18.33}}                                 & 14.77\textsuperscript{\textcolor{red}{6.87}}                                 & 27.62\textsuperscript{\textcolor{red}{7.22}}                                 & 15.50\textsuperscript{\textcolor{red}{8.46}}                                  & 27.89\textsuperscript{\textcolor{red}{10.95}}                                 & 11.75\textsuperscript{\textcolor{red}{3.56}}                                            & 27.20\textsuperscript{\textcolor{red}{7.66}}                                   \\
\bottomrule
\end{tabular}
}
\end{table*}

\paragraph{RQ3: ``Can \apifinder effectively retrieve useful APIs?''}
Prompting LLMs with oracle APIs can unlock the potential to invoke private APIs.
However, providing oracle APIs is not practical.
Therefore, whether our \apifinder can retrieve useful APIs?
Table~\ref{tab:main_results} indicates that all models with Top$K$ APIs retrieved by \apifinder perform better than those with No API setting.
This observation demonstrates that \apifinder is capable of retrieving useful APIs.
Moreover, \apifinder with human involvement generally exhibits superior performance by manually selecting potentially useful APIs. 
Surprisingly, the Top$K$ or Human setting may occasionally outperform the Oracle setting. This may stem from the noisy APIs during the training of \apicoder.

\paragraph{RQ4: ``Can \apicoder effectively invoke private APIs?''}
Table~\ref{tab:main_results} shows that almost existing models like \codegen exhibit superior performance on four private library benchmarks with prompting APIs compared to the No API setting.
Such observation demonstrates the capability of off-the-shelf \apicoder to invoke private APIs.
Although these models have achieved significant advancements, there is still room for further improvement, as indicated by the relatively low values.
In pursuit of more extraordinary performance, we develop a more advanced model named \codegenapi via continuous pre-training \codegen.
Specifically, we train a total of nine versions, three for each of the \codegen models ($350$M, $2$B, and $6$B), using three prompts as detailed in Figure~\ref{fig:apicoder}.
Table~\ref{tab:analysis_results} presents the performance of \codegenapi on four benchmarks.
We observe that \codegenapi consistently outperforms \codegen.
This indicates that \codegenapi has strengthened its capability to invoke private APIs via training on our crawled $31$ public libraries.
Note that \beatnumeval shows relatively limited gains from our approach compared to other benchmarks.
After a comprehensive analysis, we find some problems in \beatnumeval do not require invoking APIs, such as `\texttt{x[:,None]+y*8}', while our approach solely support explicitly invoked API calls, rendering it ineffective.
Overall, a vast array of experiments reveal that our \apicoder possesses the ability to invoke private APIs.

\subsection{In-Depth Study}

In this section, we will delve into a comprehensive analysis of our proposed methodology through a wide array of experiments, with the aim of providing readers with valuable insights.

\begin{figure*}
	\centering
	\begin{subfigure}[t]{0.45\linewidth}
		\centering
           \includegraphics[width=0.9\linewidth]{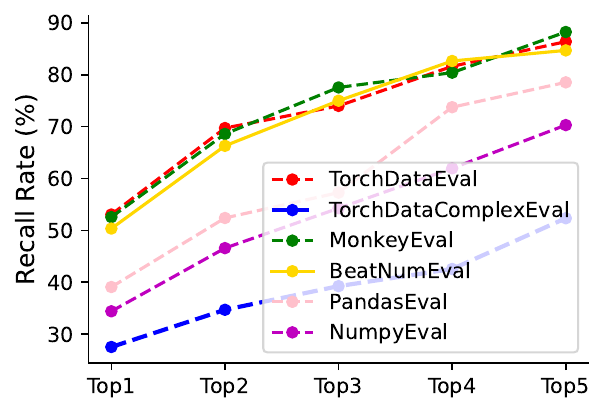}
		\caption{}
		\label{fig:recall_rate}
	\end{subfigure}
	\hfill
	\begin{subfigure}[t]{0.45\linewidth}
		\centering
		\includegraphics[width=0.9\linewidth]{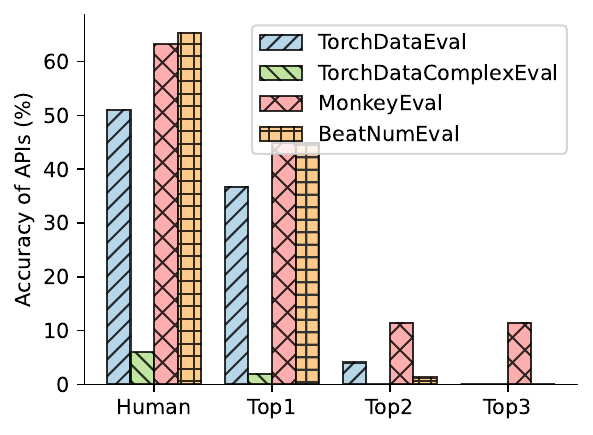}
		\caption{}
		\label{fig:accuracy}
	\end{subfigure}
	\caption{(a) The recall rate (\%) and (b) accuracy (\%) of \apifinder on private library benchmarks under various settings.}
	\label{fig:recall_accuracy_total}
\end{figure*}

\subsubsection{Quality of Retrieved APIs}
Providing high-quality APIs as prompts to \apicoder is crucial for generating private code snippets; we thus would like to analyze the quality of APIs retrieved by \apifinder or manually selected by users.
We analyze the recall rate of \apifinder on six benchmarks, and the results are displayed in Figure~\ref{fig:recall_rate}.
We observe that the recall rate of the Top$5$ surpasses $50$\% on all benchmarks. 
Therefore, it is reasonable to provide the top $5$ APIs to users during the API selection process (Figure~\ref{fig:human_in_the_loop}).
Particularly, \torchdataeval, \monkeyeval, and \beatnumeval exhibit Top$5$ recall rates close to $90$\%.
This is primarily due to their \apidoc containing a relatively small number of APIs and their programming problems being relatively simple.
As shown in Figure~\ref{fig:accuracy}, we compare the accuracy of the APIs selected by users with the APIs retrieved by \apifinder.
Here, we define accuracy as $1$ if all APIs in the programming problem are retrieved, and $0$ otherwise.
The results demonstrate that user's engagement with \apifinder can lead to a notably positive impact on accuracy.
Meanwhile, we have noticed a significantly lower accuracy of \torchdatacomplexeval compared to other benchmarks, even with human involvement, where it achieves a mere $5$\% accuracy.
Such low accuracy highlights the challenge it poses.

\begin{table}[width=1.0\linewidth,pos=h]
\centering
\caption{Performance comparison of \apifinder using dual-encoder and single-encoder on two private library benchmarks with Codex in the ``Top2/B'' setting.}
\label{tab:single_encoder}
\resizebox{1.0\linewidth}{!}{
\begin{tabular}{lllll} 
\toprule
\multirow{2}{*}{\textbf{APIFinder}} & \multicolumn{2}{c}{\textbf{TorchDataEval}} & \multicolumn{2}{c}{\begin{tabular}[c]{@{}c@{}}\textbf{TorchData}\\\textbf{ComplexEval}\end{tabular}}  \\ 
\cline{2-5}
                                    & pass@1 & pass@10                           & pass@1 & pass@10                                                                                      \\ 
\hline\hline
Dual-encoder                                & 19.06  & 49.70                             & 6.78   & 15.60                                                                                        \\
Single-encoder                              & 19.73  & 50.63                             & 6.46   & 15.98                                                                                        \\
\bottomrule
\end{tabular}
}
\end{table}

\subsubsection{Single-encoder vs. Dual-encoder}
\apifinder uses dual-encoder by default.
Technically, single-encoder can also be employed in \apifinder.
Therefore, we compare the performance of single-encoder and dual-encoder on \torchdataeval and \torchdatacomplexeval (Table~\ref{tab:single_encoder}).
We observe that the single-encoder exhibits a slight advantage over the dual-encoder in performance.
However, considering the minor performance gap and the inference speed, we ultimately opt for the dual-encoder as the default as outlined in Section~\ref{sec:apifinder}.

\begin{figure}
	\centering
	\includegraphics[scale=.75]{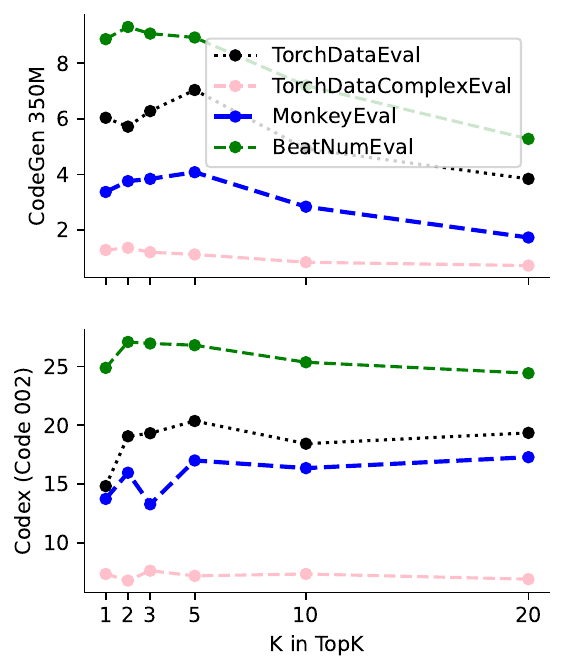}
        \caption{Pass$@1$ (\%) of \codegen 350M and Codex in various $K\in{1,2,3,5,10,20}$ in Top$K$ with prompting API basic (Top$K$/B setting).}
	\label{fig:top_k_accuracy}
\end{figure}

\subsubsection{Different $K$ in Top$K$}
As the value of $K$ in Top$K$ increases, not only does the recall rate improve, but the noise introduced to \apicoder also escalates, and vice versa. Therefore, we would like to determine the most suitable value of $K$.
In detail, we compare the pass$@1$ changes of two models with different sizes, under varying values of $K$, on four private library benchmarks.
As depicted in Figure~\ref{fig:top_k_accuracy}, it is intriguing to highlight a contrasting disparity in the sensitivities of the two models to $K$.
When $K$ exceeds $5$, the performance of the small model \codegen $350$M deteriorates across all benchmarks, while the large model Codex shows stability with no decline.
Such an observation implies that the large model exhibits superior robustness to excessive noise compared to the small one.
Overall, the choice of a suitable $K$ should take into account factors such as the model parameters, the model performance, and the intrinsic characteristics of the benchmark.


\subsubsection{Different Model Sizes}
\begin{figure}
	\centering
	\includegraphics[scale=.63]{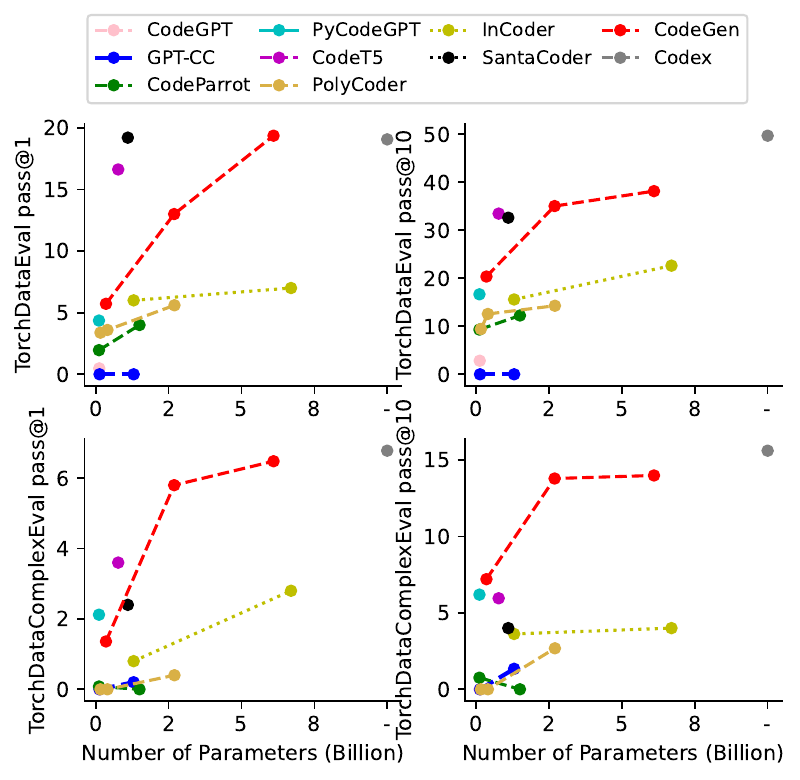}
        \caption{Parameter size vs. pass$@k$: a comparative analysis of $10$ popular models on \torchdataeval and \torchdatacomplexeval in the ``Top$2$/B'' setting.}
\label{fig:model_size_private_performance}
\end{figure}

\begin{figure}
	\centering
	\includegraphics[scale=.63]{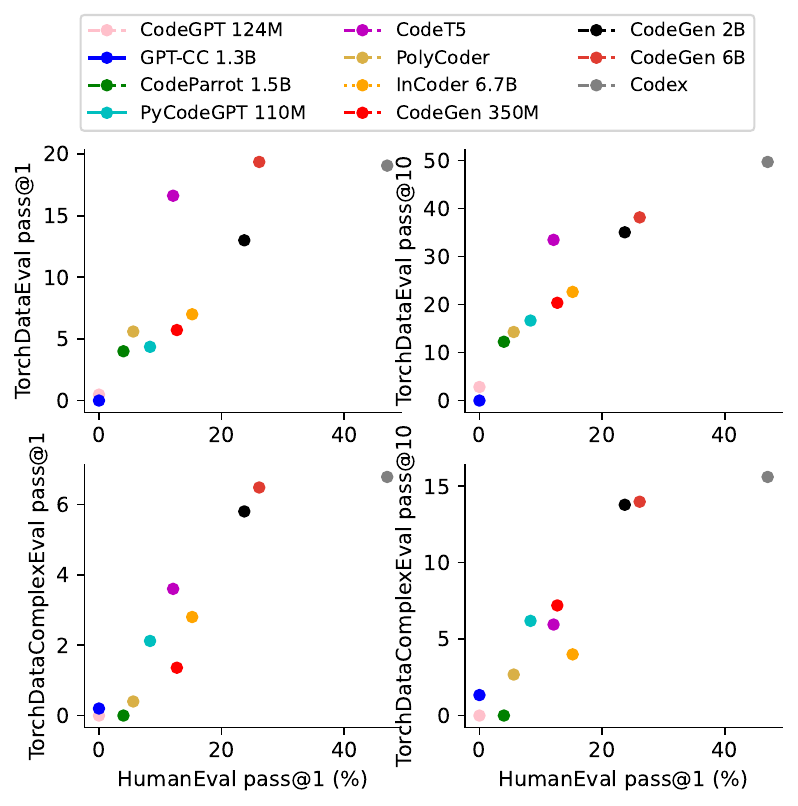}
        \caption{Performance analysis of $11$ code generation models on HumanEval~\cite{codex} and \torchdataeval (\torchdatacomplexeval) in the ``Top$2$/B'' setting.}
	\label{fig:humaneval_private_performance}
\end{figure}

It is a well-known notion that emergent ability emerges when model parameters are sufficient~\cite{emergent_ability}.
So, we would like to investigate the effect of the magnitude of parameters in \apicoder on the performance in private libraries.
Specifically, we compare pass$@1$ and pass$@10$ on \torchdataeval and \torchdatacomplexeval using $10$ models of varying sizes.
The results are depicted in Figure~\ref{fig:model_size_private_performance}.
We can find that larger models generally result in improved performance in private libraries.
Unfortunately, performance still remains at a low level, even with a considerable number of parameters.
For example, the gigantic model, Codex (code-davinci-002), achieve only a $15.60$\% pass$@10$ on \torchdatacomplexeval.
This phenomenon also confirms the formidable challenge of private-library-oriented code generation.
Furthermore, we analyze the performance correlation of $11$ code generation models on both HumanEval~\cite{codex} and private libraries in Figure~\ref{fig:humaneval_private_performance}, where HumanEval is currently the most popular benchmark for evaluating code generation capabilities.
We observe that models that perform well on HumanEval also stand out in the private library benchmarks.
Hence, we believe that our proposed private library scenario stands to benefit from the rapid progress in general code generation techniques.

\begin{table*}[width=2.0\linewidth,pos=h]
\centering
\caption{Ablation study of \codegenapi 350M trained with ``prompt\#1'' in the ``Human/B'' setting. The default noise rate is $5$\%.}
\label{tab:noise_resample_results}
\resizebox{0.88\linewidth}{!}{
\begin{tabular}{lllllllll} 
\toprule
\multirow{2}{*}{\textbf{APICoder}} & \multicolumn{2}{c}{\textbf{TorchDataEval}} & \multicolumn{2}{c}{\begin{tabular}[c]{@{}c@{}}\textbf{TorchData}\\\textbf{ComplexEval}\end{tabular}} & \multicolumn{2}{c}{\textbf{MonkeyEval}} & \multicolumn{2}{c}{\textbf{BeatNumEval}}  \\ 
\cline{2-9}
                                   & pass@1         & pass@10                   & pass@1        & pass@10                                                                              & pass@1        & pass@10                 & pass@1         & pass@10                  \\ 
\hline\hline
CodeGenAPI 350M                    & \textbf{14.97} & \textbf{35.86}            & \textbf{2.58} & \textbf{8.75}                                                                        & \textbf{6.30} & \textbf{16.45}          & \textbf{12.92} & \textbf{33.56}           \\ 
\hdashline
~- w/ noise rate 0\%               & 12.14          & 33.52                     & 2.20          & 4.24                                                                                 & 5.12          & 16.18                   & 12.89          & 31.44                    \\
~- w/ noise rate 10\%              & 14.05          & 34.13                     & 2.53          & 6.57                                                                                 & 5.06          & 15.76                   & 12.21          & 33.36                    \\
~- w/ noise rate 20\%              & 12.46          & 32.74                     & 2.25          & 5.61                                                                                 & 5.17          & 14.36                   & 11.00          & 31.32                    \\
~- w/o resampling                  & 12.27          & 31.57                     & 2.36          & 7.14                                                                                 & 5.69          & 15.67                   & 11.26          & 32.90                    \\
\bottomrule
\end{tabular}
}
\end{table*}

\subsubsection{Noise Rate}
A carefully chosen noise rate is crucial for \codegenapi to handle a diverse range of APIs.
If the noise rate is excessively high, it will disrupt the original distribution; conversely, it will lose the ability to address noise APIs if too low.
We thus aim to explore the impact of noise rate on \apicoder.
The default noise rate for \codegenapi is $5$\%, and we also experiment with $0$\%, $10$\%, and $20$\% in Table~\ref{tab:noise_resample_results}.
The results indicate that $5$\% is the optimal choice, with too little or excessive noise APIs causing a decrease in performance.

\subsubsection{Re-sampling strategy}
During the continuous pre-training of \codegenapi, we employ a re-sampling strategy.
As mentioned in Section~\ref{sec:apicoder}, the core idea of this strategy is to make high-quality Python files more readily trainable and vice versa.
To validate the effectiveness of this strategy, we omit it during the training of \codegenapi $350$M, as shown in Table~\ref{tab:noise_resample_results}.
The results show a sustained decline in performance, demonstrating the validity of the re-sampling strategy.

\begin{figure}
	\centering
	\includegraphics[scale=.55]{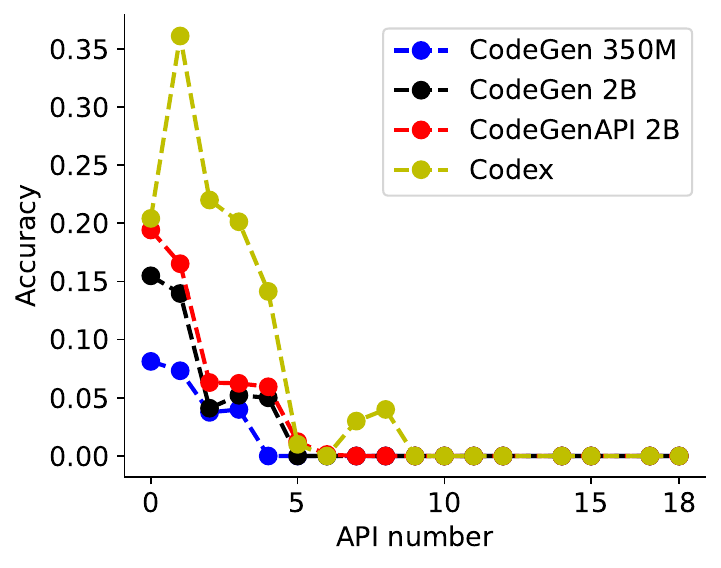}
        \caption{API number vs. accuracy on all private library benchmarks in the ``Oracle/B'' setting.}
	\label{fig:api_num_accuracy}
\end{figure}

\subsubsection{Different Difficulty}
\apicoder has the capability to tackle programming problems in private libraries.
We are curious about what level of difficulty \apicoder can solve in private library programming problems.
Hence, we aim to access \apicoder's performance with problems of varying difficulty through accuracy evaluations under varying API counts.
To be specific, we calculate the accuracy of four models across varying API counts on a combined set of our released four benchmarks.
The results are illustrated in Figure~\ref{fig:api_num_accuracy}.
Our finding is that larger models possess a heightened ability to solve complex problems.
For instance, Codex even resolves these programming problems that include $8$ APIs.
Meanwhile, we also observe that \codegenapi consistently outperforms \codegen across varying API counts.
For example, \codegenapi $2$B is capable of resolving these programming problems that include $5$ APIs, while \codegen $2$B is not.
Such an observation demonstrates our \codegenapi continuously pre-training on public libraries does enhance its ability to call private APIs.

\begin{figure}
	\centering
	\includegraphics[scale=.7]{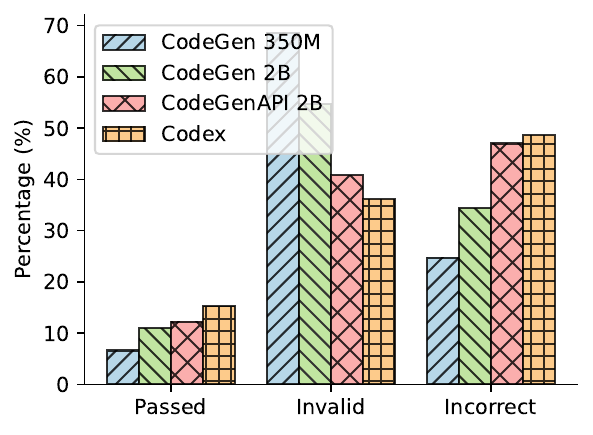}
        \caption{Comparative analysis of passed, invalid, and incorrect API usage proportions for four models in \torchdataeval with `Oracle/B' setting. `Invalid' refers to when the model does not invoke the prompted API, while `Incorrect' denotes instances where the API is invoked, but not used properly.}
	\label{fig:error_analysis}
\end{figure}

\subsubsection{Error Type}
When provided with Oracle APIs, \apicoder can solve some private library problems, but a substantial portion remains unresolved.
We are intrigued by the reasons behind these unresolved problems, whether they stem from the lack of API calls (labeled as Invalid) or incorrect API usage (labeled as Incorrect).
Thus, we compare the passed, invalid, and incorrect rates of four models using \torchdataeval in Figure~\ref{fig:error_analysis}.
Our finding is that the models with superior performance exhibit higher passed and incorrect rates and lower invalid rates.
This highlights that, on the one hand, superior-performing models generate candidate code more readily passing test cases. On the other hand, the majority of errors in these superior models are due to incorrect API usage, whereas subpar models fail to even invoke APIs.
Our another finding is that \codegenapi $2$B outperforms \codegen $2$B in both the passed and the incorrect rates while maintaining lower invalid rates.
This also obliquely reflects that the continuous pre-training from \codegen to \codegenapi indeed enhances the ability to invoke private APIs.

\begin{table}[width=1.0\linewidth,pos=h]
\centering
\caption{Pass$@k$ (\%) of \codegen 2B, \codegenapi 2B and Codex on two public library benchmarks. \#1, \#2, and \#3 represent \codegenapi trained with the three prompts in Figure~\ref{fig:apicoder}. The values of \textcolor{red}{red} and \textcolor{darkgreen}{green} represent improvements and deteriorations compared to the No API setting.}
\label{tab:public_results}
\resizebox{1.0\linewidth}{!}{
\begin{tabular}{c:c:l|llll} 
\toprule
\multicolumn{1}{l}{\multirow{3}{*}{\textbf{APICoder}}}                      & \multicolumn{1}{l}{\multirow{3}{*}{\textbf{P.}}} & \multicolumn{1}{l}{\multirow{3}{*}{\textbf{APIFinder}}} & \multicolumn{2}{c}{\textbf{PandasEval}}                                                                                                                                 & \multicolumn{2}{c}{\textbf{NumpyEval}}                                                                                                                                   \\ 
\cline{4-7}
\multicolumn{1}{l}{}                                                        & \multicolumn{1}{l}{}                             & \multicolumn{1}{l}{}                                    & \multicolumn{4}{c}{\passk}                                                                                                                                                                                                                                                                                                                         \\
\multicolumn{1}{l}{}                                                        & \multicolumn{1}{l}{}                             & \multicolumn{1}{l}{}                                    & k=1                                                                                & k=10                                                                               & k=1                                                                                & k=10                                                                                \\ 
\hline\hline
\multirow{10}{*}{\begin{tabular}[c]{@{}c@{}}CodeGen\\2B\end{tabular}}       & \multirow{10}{*}{-}                              & No API                                                  & 37                                                                                 & 63                                                                                 & 38                                                                                 & 66                                                                                  \\
                                                                            &                                                  & {\cellcolor[rgb]{0.788,0.918,1}}Oracle/B                & {\cellcolor[rgb]{0.788,0.918,1}}32\textsuperscript{\textcolor[rgb]{0,0.502,0}{5}}  & {\cellcolor[rgb]{0.788,0.918,1}}56\textsuperscript{\textcolor[rgb]{0,0.502,0}{7}}  & {\cellcolor[rgb]{0.788,0.918,1}}36\textsuperscript{\textcolor[rgb]{0,0.502,0}{2}}  & {\cellcolor[rgb]{0.788,0.918,1}}61\textsuperscript{\textcolor[rgb]{0,0.502,0}{5}}   \\
                                                                            &                                                  & {\cellcolor[rgb]{0.788,0.918,1}}Oracle/E                & {\cellcolor[rgb]{0.788,0.918,1}}29\textsuperscript{\textcolor[rgb]{0,0.502,0}{8}}  & {\cellcolor[rgb]{0.788,0.918,1}}49\textsuperscript{\textcolor[rgb]{0,0.502,0}{14}} & {\cellcolor[rgb]{0.788,0.918,1}}31\textsuperscript{\textcolor[rgb]{0,0.502,0}{7}}  & {\cellcolor[rgb]{0.788,0.918,1}}49\textsuperscript{\textcolor[rgb]{0,0.502,0}{17}}  \\
                                                                            &                                                  & {\cellcolor[rgb]{0.788,0.918,1}}Oracle/BE               & {\cellcolor[rgb]{0.788,0.918,1}}25\textsuperscript{\textcolor[rgb]{0,0.502,0}{12}} & {\cellcolor[rgb]{0.788,0.918,1}}50\textsuperscript{\textcolor[rgb]{0,0.502,0}{13}} & {\cellcolor[rgb]{0.788,0.918,1}}27\textsuperscript{\textcolor[rgb]{0,0.502,0}{11}} & {\cellcolor[rgb]{0.788,0.918,1}}47\textsuperscript{\textcolor[rgb]{0,0.502,0}{19}}  \\
                                                                            &                                                  & Top1/B                                                  & 28\textsuperscript{\textcolor[rgb]{0,0.502,0}{9}}                                  & 55\textsuperscript{\textcolor[rgb]{0,0.502,0}{8}}                                  & 33\textsuperscript{\textcolor[rgb]{0,0.502,0}{5}}                                  & 53\textsuperscript{\textcolor[rgb]{0,0.502,0}{13}}                                  \\
                                                                            &                                                  & Top2/B                                                  & 27\textsuperscript{\textcolor[rgb]{0,0.502,0}{10}}                                 & 54\textsuperscript{\textcolor[rgb]{0,0.502,0}{9}}                                  & 32\textsuperscript{\textcolor[rgb]{0,0.502,0}{6}}                                  & 56\textsuperscript{\textcolor[rgb]{0,0.502,0}{10}}                                  \\
                                                                            &                                                  & Top2/E                                                  & 23\textsuperscript{\textcolor[rgb]{0,0.502,0}{14}}                                 & 45\textsuperscript{\textcolor[rgb]{0,0.502,0}{18}}                                 & 22\textsuperscript{\textcolor[rgb]{0,0.502,0}{16}}                                 & 39\textsuperscript{\textcolor[rgb]{0,0.502,0}{27}}                                  \\
                                                                            &                                                  & Top2/BE                                                 & 19\textsuperscript{\textcolor[rgb]{0,0.502,0}{18}}                                 & 37\textsuperscript{\textcolor[rgb]{0,0.502,0}{26}}                                 & 15\textsuperscript{\textcolor[rgb]{0,0.502,0}{23}}                                 & 31\textsuperscript{\textcolor[rgb]{0,0.502,0}{35}}                                  \\
                                                                            &                                                  & Top3/B                                                  & 28\textsuperscript{\textcolor[rgb]{0,0.502,0}{9}}                                  & 56\textsuperscript{\textcolor[rgb]{0,0.502,0}{7}}                                  & 33\textsuperscript{\textcolor[rgb]{0,0.502,0}{5}}                                  & 57\textsuperscript{\textcolor[rgb]{0,0.502,0}{9}}                                   \\
                                                                            &                                                  & Top5/B                                                  & 30\textsuperscript{\textcolor[rgb]{0,0.502,0}{7}}                                  & 57\textsuperscript{\textcolor[rgb]{0,0.502,0}{6}}                                  & 31\textsuperscript{\textcolor[rgb]{0,0.502,0}{7}}                                  & 55\textsuperscript{\textcolor[rgb]{0,0.502,0}{11}}                                  \\ 
\hdashline
\multirow{9}{*}{\begin{tabular}[c]{@{}c@{}}CodeGenAPI\\2B\end{tabular}}     & \multicolumn{1}{l:}{\multirow{5}{*}{\#1}}        & {\cellcolor[rgb]{0.788,0.918,1}}Oracle/B                & {\cellcolor[rgb]{0.788,0.918,1}}41\textsuperscript{\textcolor{red}{4}}             & {\cellcolor[rgb]{0.788,0.918,1}}68\textsuperscript{\textcolor{red}{5}}             & {\cellcolor[rgb]{0.788,0.918,1}}44\textsuperscript{\textcolor{red}{6}}             & {\cellcolor[rgb]{0.788,0.918,1}}69\textsuperscript{\textcolor{red}{3}}              \\
                                                                            & \multicolumn{1}{l:}{}                            & Top1/B                                                  & 38\textsuperscript{\textcolor{red}{1}}                                             & 65\textsuperscript{\textcolor{red}{2}}                                             & 40\textsuperscript{\textcolor{red}{2}}                                             & 66\textsuperscript{\textcolor{red}{0}}                                              \\
                                                                            & \multicolumn{1}{l:}{}                            & Top2/B                                                  & 39\textsuperscript{\textcolor{red}{2}}                                             & 65\textsuperscript{\textcolor{red}{2}}                                             & 41\textsuperscript{\textcolor{red}{3}}                                             & 68\textsuperscript{\textcolor{red}{2}}                                              \\
                                                                            & \multicolumn{1}{l:}{}                            & Top3/B                                                  & 38\textsuperscript{\textcolor{red}{1}}                                             & 64\textsuperscript{\textcolor{red}{1}}                                             & 41\textsuperscript{\textcolor{red}{3}}                                             & 67\textsuperscript{\textcolor{red}{1}}                                              \\
                                                                            & \multicolumn{1}{l:}{}                            & Top5/B                                                  & 37\textsuperscript{\textcolor{red}{0}}                                             & 64\textsuperscript{\textcolor{red}{1}}                                             & 40\textsuperscript{\textcolor{red}{2}}                                             & 68\textsuperscript{\textcolor{red}{2}}                                              \\ 
\cdashline{2-2}
                                                                            & \multicolumn{1}{l:}{\multirow{2}{*}{\#2}}        & {\cellcolor[rgb]{0.788,0.918,1}}Oracle/E                & {\cellcolor[rgb]{0.788,0.918,1}}42\textsuperscript{\textcolor{red}{5}}             & {\cellcolor[rgb]{0.788,0.918,1}}66\textsuperscript{\textcolor{red}{3}}             & {\cellcolor[rgb]{0.788,0.918,1}}44\textsuperscript{\textcolor{red}{6}}             & {\cellcolor[rgb]{0.788,0.918,1}}71\textsuperscript{\textcolor{red}{5}}              \\
                                                                            & \multicolumn{1}{l:}{}                            & Top2/E                                                  & 37\textsuperscript{\textcolor{red}{0}}                                             & 62\textsuperscript{\textcolor[rgb]{0,0.502,0}{1}}                                  & 34\textsuperscript{\textcolor[rgb]{0,0.502,0}{4}}                                  & 61\textsuperscript{\textcolor[rgb]{0,0.502,0}{5}}                                   \\ 
\cdashline{2-2}
                                                                            & \multicolumn{1}{l:}{\multirow{2}{*}{\#3}}        & {\cellcolor[rgb]{0.788,0.918,1}}Oracle/BE               & {\cellcolor[rgb]{0.788,0.918,1}}41\textsuperscript{\textcolor{red}{4}}             & {\cellcolor[rgb]{0.788,0.918,1}}66\textsuperscript{\textcolor{red}{3}}             & {\cellcolor[rgb]{0.788,0.918,1}}45\textsuperscript{\textcolor{red}{7}}             & {\cellcolor[rgb]{0.788,0.918,1}}70\textsuperscript{\textcolor{red}{4}}              \\
                                                                            & \multicolumn{1}{l:}{}                            & Top2/BE                                                 & 35\textsuperscript{\textcolor[rgb]{0,0.502,0}{2}}                                  & 61\textsuperscript{\textcolor[rgb]{0,0.502,0}{2}}                                  & 35\textsuperscript{\textcolor[rgb]{0,0.502,0}{3}}                                  & 58\textsuperscript{\textcolor[rgb]{0,0.502,0}{8}}                                   \\ 
\hline
\multirow{10}{*}{\begin{tabular}[c]{@{}c@{}}Codex\\(Code 002)\end{tabular}} & \multirow{10}{*}{-}                              & No API                                                  & 54                                                                                 & 83                                                                                 & 62                                                                                 & 91                                                                                  \\
                                                                            &                                                  & {\cellcolor[rgb]{0.788,0.918,1}}Oracle/B                & {\cellcolor[rgb]{0.788,0.918,1}}52\textsuperscript{\textcolor[rgb]{0,0.502,0}{2}}  & {\cellcolor[rgb]{0.788,0.918,1}}80\textsuperscript{\textcolor[rgb]{0,0.502,0}{3}}  & {\cellcolor[rgb]{0.788,0.918,1}}61\textsuperscript{\textcolor[rgb]{0,0.502,0}{1}}  & {\cellcolor[rgb]{0.788,0.918,1}}87\textsuperscript{\textcolor[rgb]{0,0.502,0}{4}}   \\
                                                                            &                                                  & {\cellcolor[rgb]{0.788,0.918,1}}Oracle/E                & {\cellcolor[rgb]{0.788,0.918,1}}51\textsuperscript{\textcolor[rgb]{0,0.502,0}{3}}  & {\cellcolor[rgb]{0.788,0.918,1}}81\textsuperscript{\textcolor[rgb]{0,0.502,0}{2}}  & {\cellcolor[rgb]{0.788,0.918,1}}61\textsuperscript{\textcolor[rgb]{0,0.502,0}{1}}  & {\cellcolor[rgb]{0.788,0.918,1}}88\textsuperscript{\textcolor[rgb]{0,0.502,0}{3}}   \\
                                                                            &                                                  & {\cellcolor[rgb]{0.788,0.918,1}}Oracle/BE               & {\cellcolor[rgb]{0.788,0.918,1}}50\textsuperscript{\textcolor[rgb]{0,0.502,0}{4}}  & {\cellcolor[rgb]{0.788,0.918,1}}80\textsuperscript{\textcolor[rgb]{0,0.502,0}{3}}  & {\cellcolor[rgb]{0.788,0.918,1}}60\textsuperscript{\textcolor[rgb]{0,0.502,0}{2}}  & {\cellcolor[rgb]{0.788,0.918,1}}88\textsuperscript{\textcolor[rgb]{0,0.502,0}{3}}   \\
                                                                            &                                                  & Top1/B                                                  & 47\textsuperscript{\textcolor[rgb]{0,0.502,0}{7}}                                  & 77\textsuperscript{\textcolor[rgb]{0,0.502,0}{6}}                                  & 56\textsuperscript{\textcolor[rgb]{0,0.502,0}{6}}                                  & 85\textsuperscript{\textcolor[rgb]{0,0.502,0}{6}}                                   \\
                                                                            &                                                  & Top2/B                                                  & 48\textsuperscript{\textcolor[rgb]{0,0.502,0}{6}}                                  & 81\textsuperscript{\textcolor[rgb]{0,0.502,0}{2}}                                  & 57\textsuperscript{\textcolor[rgb]{0,0.502,0}{5}}                                  & 86\textsuperscript{\textcolor[rgb]{0,0.502,0}{5}}                                   \\
                                                                            &                                                  & Top2/E                                                  & 48\textsuperscript{\textcolor[rgb]{0,0.502,0}{11}}                                 & 82\textsuperscript{\textcolor[rgb]{0,0.502,0}{1}}                                  & 53\textsuperscript{\textcolor[rgb]{0,0.502,0}{9}}                                  & 85\textsuperscript{\textcolor[rgb]{0,0.502,0}{6}}                                   \\
                                                                            &                                                  & Top2/BE                                                 & 45\textsuperscript{\textcolor[rgb]{0,0.502,0}{14}}                                 & 78\textsuperscript{\textcolor[rgb]{0,0.502,0}{5}}                                  & 49\textsuperscript{\textcolor[rgb]{0,0.502,0}{13}}                                 & 82\textsuperscript{\textcolor[rgb]{0,0.502,0}{9}}                                   \\
                                                                            &                                                  & Top3/B                                                  & 46\textsuperscript{\textcolor[rgb]{0,0.502,0}{8}}                                  & 80\textsuperscript{\textcolor[rgb]{0,0.502,0}{3}}                                  & 52\textsuperscript{\textcolor[rgb]{0,0.502,0}{10}}                                 & 85\textsuperscript{\textcolor[rgb]{0,0.502,0}{6}}                                   \\
                                                                            &                                                  & Top5/B                                                  & 45\textsuperscript{\textcolor[rgb]{0,0.502,0}{9}}                                  & 80\textsuperscript{\textcolor[rgb]{0,0.502,0}{3}}                                  & 53\textsuperscript{\textcolor[rgb]{0,0.502,0}{9}}                                  & 85\textsuperscript{\textcolor[rgb]{0,0.502,0}{6}}                                   \\
\bottomrule
\end{tabular}
}
\end{table}

\subsubsection{Public Library}
Technically, our proposed approach can be applied to public scenarios as well.
As such, Table~\ref{tab:public_results} showcases the performance of three models on \pandaseval and \numpyeval.
The results indicate a decline in performance when prompting off-the-shelf models with public APIs, such as \codegen $2$B and Codex.
A possible reason is that prompting the previously seen public APIs may disrupt the probability prediction of the model.
Also, the decline is more pronounced when the model size is smaller.
For example, \codegen $2$B see a $35$\% pass$@10$ decrease on \numpyeval in the ``Top$2$/BE'' setting.
Surprisingly, \codegenapi $2$B brings a performance gain.
For instance, \codegenapi $2$B experiences a roughly $6$\% increase in pass@1 on \numpyeval in the ``Oracle/B'' setting.
This once again highlights that our \codegenapi possesses the ability to invoke the prompted APIs effectively.

\section{Discussion and Limitations}
In this section, we will delve into some thought-provoking discussions on the limitations of our paper.
(1)
As mentioned in Section~\ref{sec:benchmark_construction}, creating a truly private library benchmark poses a significant challenge.
As a result, in addition to \torchdataeval and \torchdatacomplexeval, we also derive two pseudo private library benchmarks from public ones.
Despite our best efforts to modify public libraries into private ones by paraphrasing keywords and \apidoc, there remains a potential risk to the validity and fairness of the private library evaluation.
So, it is a worthwhile endeavor to gather more real-world private libraries and corresponding programming tasks in future work.
(2)
As outlined in Section~\ref{sec:main_results}, our proposed approach yields superior performance on these models with larger parameters.
Likewise, if the model has fewer parameters, the benefits yielded by our approach may be limited or even ineffective.
As a result, our approach is relatively sensitive to the capabilities of the base model itself.
(3)
As early explorers in the field of private-library-oriented code generation, our constructed private libraries typically feature a relatively modest API count (about $200$).
In this case, \apifinder can retrieve some useful APIs.
However, as the API count grows, the challenge faced by \apifinder could become amplified.
Even in our benchmarks with relatively fewer APIs, the performance of \apifinder lag significantly behind the ``Oracle'' setting (Table~\ref{tab:main_results}).
This reveals the ample room for improvement in \apifinder.
(4)
Providing API examples to \apicoder may introduce a minor bias in evaluating \torchdataeval, as the construction of \torchdataeval also refers to API examples in \apidoc.
(5)
Unavoidably, our paper entails a heavy consumption of computational resources.
Therefore, we will publicly release the LLMs-generated files to foster further research.
(6)
Our approach focuses solely on Python. When extrapolated to other programming languages, some potential threats may exist due to subtle differences between them.
(7)
Several powerful code generation models, such as PaLM-Coder~\cite{palm}, PanGu-Coder~\cite{pangu-coder}, and AlphaCode~\cite{alphacode}, are not publicly available, which prevents us from including them in our experiments. In light of this, we have made every effort to run all accessible models listed in Table~\ref{tab:main_results}, aiming to obtain the most trustworthy results possible.
(8)
Practically speaking, one intriguing idea is to convert our approach into a programming assistant to aid developers in better using private libraries, as private libraries are a common occurrence in routine coding scenarios. Considering potential privacy and security concerns, this remains a topic for future research.

\section{Conclusion}
In this paper, we propose a novel scenario for code generation focused on private libraries.
To address this scenario, we design a framework by simulating the human use of private libraries, which consists of two modules: \apifinder and \apicoder.
\apifinder first retrieves relevant APIs from \apidoc, and \apicoder then utilizes these APIs to solve programming problems.
Additionally, we craft four private library benchmarks, including \torchdataeval, \torchdatacomplexeval, \monkeyeval, and \beatnumeval.
Lastly, we carry out extensive experiments on the four benchmarks, showcasing the strengths and limitations of our approach, which could provide some meaningful insights for future work.
Moving forward, our goal is to encapsulate our approach as an auxiliary tool designed to facilitate the coding process for programmers. This aim presents several challenging yet promising research questions.
For instance, how to ensure privacy and security when employing LLMs? How to tackle the scenario of mixed usage of public and private libraries? How can we measure and increase the trust that developers place in the generated code? How to design an effective user interface that allows programmers to interact smoothly with the tool? And beyond these, there are countless other intriguing questions in this field that await exploration.

\appendix

\section{Keyword Conversion from Public to Private Libraries} \label{apx:keyword_conversion}
We manually convert the public libraries into private ones by paraphrasing all relevant keywords in Section~\ref{sec:benchmark_construction}. Table~\ref{tab:public_to_private_conversion} lists all the keywords before and after converting \pandaseval(\numpyeval) to \monkeyeval(\beatnumeval).

\begin{table*}[width=2.0\linewidth,pos=h]
\centering
\caption{Keywords conversion from \pandaseval (\numpyeval) to \monkeyeval (\beatnumeval).}
\label{tab:public_to_private_conversion}
\resizebox{1.0\linewidth}{!}{
\begin{tabular}{lllllll} 
\toprule
\multicolumn{7}{c}{\textbf{PandasEval - MonkeyEval}}                                                                                                                                                   \\ 
\hline
\rowcolor[rgb]{0.843,0.843,0.843} df           & Pandas        & pandas                                      & len           & tolist          & isin               & sort\_index                      \\
kf                                             & Monkey        & monkey                                      & length        & convert\_list   & incontain          & sorting\_index                   \\
\rowcolor[rgb]{0.843,0.843,0.843} isnull       & apply         & to\_numeric                                 & dropna        & append          & tail               & value\_counts                    \\
ifnull                                         & employ        & to\_num                                     & sipna         & adding          & last\_tail         & counts\_value\_num               \\
\rowcolor[rgb]{0.843,0.843,0.843} innull       & astype        & select\_dtypes                              & iterrows      & min             & max                & drop\_duplicates                 \\
isnone                                         & totype        & choose\_dtypes                              & traversal     & get\_min        & get\_max           & remove\_duplicates               \\
\rowcolor[rgb]{0.843,0.843,0.843} pd           & shift         & merge                                       & copy          & rename\_axis    & reset\_index       & sample                           \\
mk                                             & shifting      & unioner                                     & clone         & renaming\_axis  & reseting\_index    & sample\_by\_num                  \\
\rowcolor[rgb]{0.843,0.843,0.843} concat       & to\_dict      & cumsum                                      & last          & to\_string      & applymap           & duplicated                       \\
concating                                      & convert\_dict & cumulative\_sum                             & final\_item   & convert\_string & conduct\_map       & duplicated\_values               \\
\rowcolor[rgb]{0.843,0.843,0.843} isna         & format        & div                                         & mean          & ceil            & assign             & DataFrame                        \\
ifna                                           & formating     & division                                    & average       & ceiling         & allocate           & KnowledgeFrame                   \\
\rowcolor[rgb]{0.843,0.843,0.843} drop         & Series        & ravel                                       & any           & fillna          & all                & to\_pydatetime                   \\
sip                                            & Collections   & flat\_underlying                            & whatever      & fillnone        & total\_all         & convert\_pydatetime              \\
\rowcolor[rgb]{0.843,0.843,0.843} reindex      & head          & sort\_values                                & rename        & sum             & unique             & to\_datetime                     \\
reindexing                                     & header\_num   & sort\_the\_values                           & renaming      & total\_sum      & distinctive        & convert\_datetime                \\
\rowcolor[rgb]{0.843,0.843,0.843} map          & std           & intersection                                & groupby       & nlargest        & replace            & dataframe                        \\
mapping                                        & standard      & \textcolor[rgb]{0.039,0.188,0.412}{interst} & grouper       & nbiggest        & replacing          & knowledgeframe                   \\
\rowcolor[rgb]{0.843,0.843,0.843} get          & series        & round                                       &               &                 &                    &                                  \\
getting                                        & collections   & value\_round                                &               &                 &                    &                                  \\ 
\hline
\multicolumn{7}{c}{\textbf{NumpyEval - BeatNumEval}}                                                                                                                                                   \\ 
\hline
\rowcolor[rgb]{0.843,0.843,0.843} np           & Numpy         & array                                       & unique        & ndarray         & transpose          & reshape                          \\
bn                                             & Beatnum       & numset                                      & uniq          & ndnumset        & switching\_places  & change\_shape\_to                \\
\rowcolor[rgb]{0.843,0.843,0.843} real         & numpy         & vstack                                      & sum           & imag            & in1d               & flatten                          \\
reality                                        & beatnum       & vertical\_stack                             & total\_count  & imaginary       & intersection1dim   & convert\_into\_one\_dim          \\
\rowcolor[rgb]{0.843,0.843,0.843} isnan        & all           & fromstring                                  & inv           & mean            & where              & compressed                       \\
ifnan                                          & total         & come\_from\_str                             & inverse       & average         & filter\_condition  & remove\_masked\_data             \\
\rowcolor[rgb]{0.843,0.843,0.843} add          & max           & histogram                                   & to\_numpy     & filled          & stack              & cumsum                           \\
add\_concat                                    & get\_max      & hist\_operation                             & to\_beatnum   & masked\_fill    & pile\_operation    & cumulative\_sum                  \\
\rowcolor[rgb]{0.843,0.843,0.843} insert       & arange        & ravel                                       & std           & argmax          & argmin             & full                             \\
stick                                          & arr\_range    & asview                                      & standard\_op  & get\_argmax     & get\_argmin\_value & full\_value\_func                \\
\rowcolor[rgb]{0.843,0.843,0.843} slice        & squeeze       & hstack                                      & asarray       & repeat          & bincount           & unravel\_index                   \\
piece                                          & sqz           & horizontal\_stack                           & asnumset      & duplicate       & binoccurrence      & convert\_index\_or\_arr          \\
\rowcolor[rgb]{0.843,0.843,0.843} diff         & concatenate   & any                                         & column\_stack & norm            & delete             & logical\_and                     \\
difference                                     & connect       & any\_condition                              & stack\_col    & normlizattion   & remove\_operation  & logic\_and\_element\_wise        \\
\rowcolor[rgb]{0.843,0.843,0.843} append       & split         & ones                                        & vectorize     & fill\_diagonal  & argpartition       & setxor1d                         \\
apd                                            & sep\_split    & create\_ones                                & vectorisation & pad\_diagonal   & perform\_partition & seting\_exclusive\_or\_one\_dim  \\
\rowcolor[rgb]{0.843,0.843,0.843} array\_split & abs           & astype                                      & searchsorted  & min             & fromarrays         &                                  \\
split\_array                                   & absolute      & convert\_type                               & find\_sorted  & get\_min        & come\_from\_arrays &                                  \\
\bottomrule
\end{tabular}
}
\end{table*}

\printcredits

\bibliographystyle{cas-model2-names}

\bibliography{cas-refs}



\vskip500pt

\bio{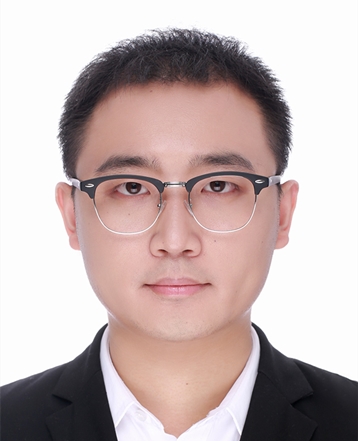}
\textbf{Daoguang Zan} is currently pursuing the Ph.D. degree with the Institute of Software, Chinese Academy of Sciences, Beijing, China. 
His principal research interest includes natural language processing and software engineering, especially in code generation and large language model. In these areas, he has published around 10 papers in top-tier conference proceedings, including ACL, IJCAI, EMNLP, ICLR, PAKDD, etc.
\endbio


\bio{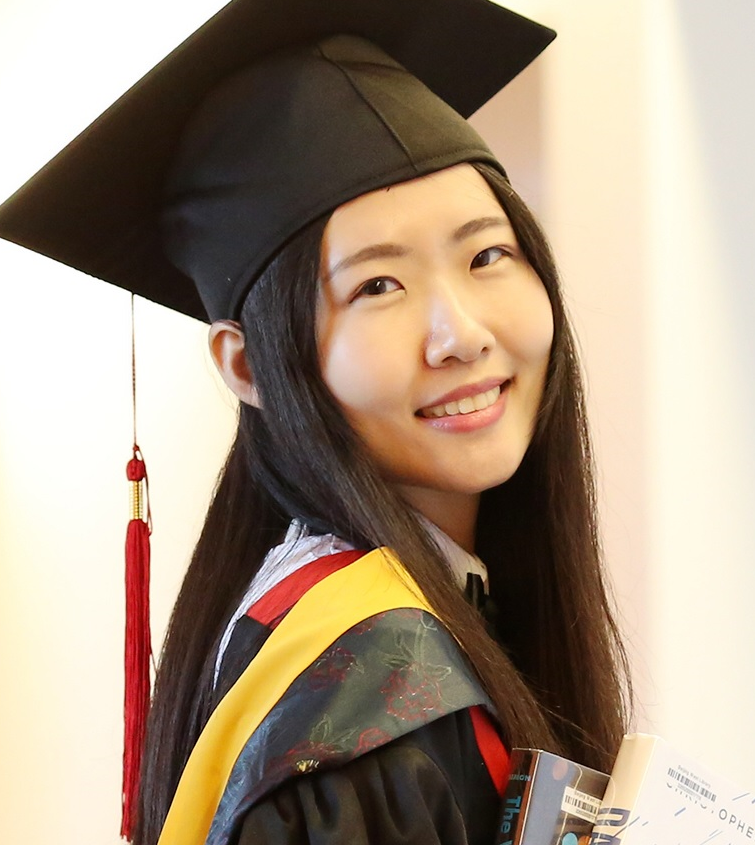}
\textbf{Bei Chen} is currently a senior researcher at Microsoft. She received her Ph.D. degree from the Department of Computer Science and Technology at Tsinghua University, Beijing, China, in 2017. She is mainly working on natural language processing, including semantic parsing, dialogue systems, pre-trained language models, and their applications in code intelligence. She has published above 30 papers in top conferences, including ICLR, NeurIPS, ACL, EMNLP, KDD, AAAI, IJCAI, etc.
\endbio


\bio{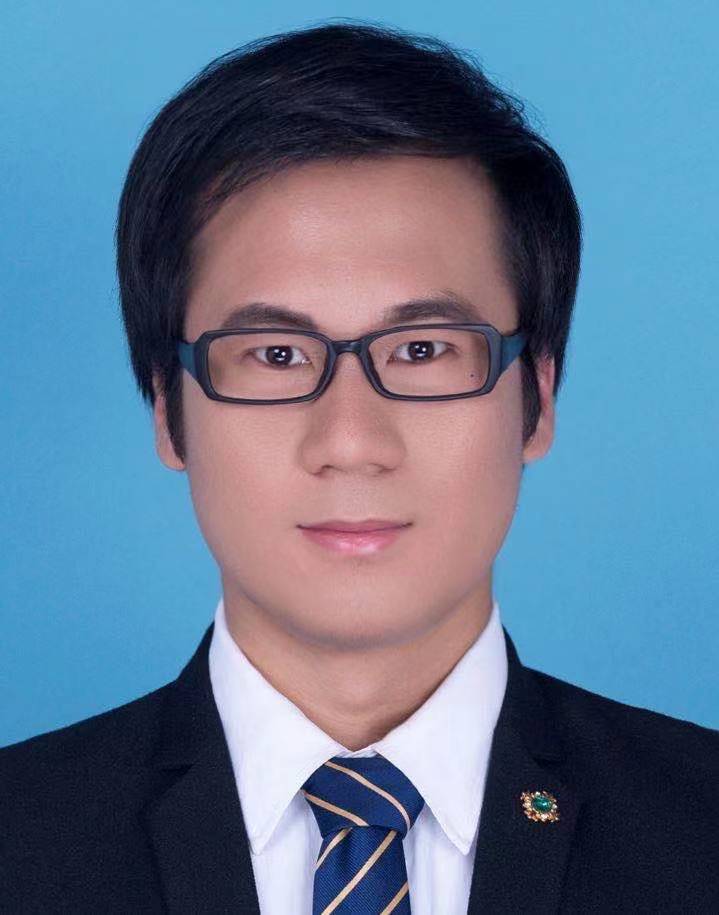}
\textbf{Yongshun Gong} is an Associate Professor at School of Software, Shandong University, China. He received his Ph.D. degree from University of Technology Sydney in 2021. His principal research interest covers the data science and machine learning, in particular, the following areas: adaptive model; spatiotemporal data mining; traffic prediction; recommender system and sequential pattern mining. He has published above 40 papers in top journals and refereed conference proceedings, including the IEEE T-PAMI, IEEE T-KDE, IEEE T-NNLS, IEEE T-CYB, IEEE T-MM, NeurIPS, CVPR, KDD, CIKM, AAAI, IJCAI, etc.
\endbio


\bio{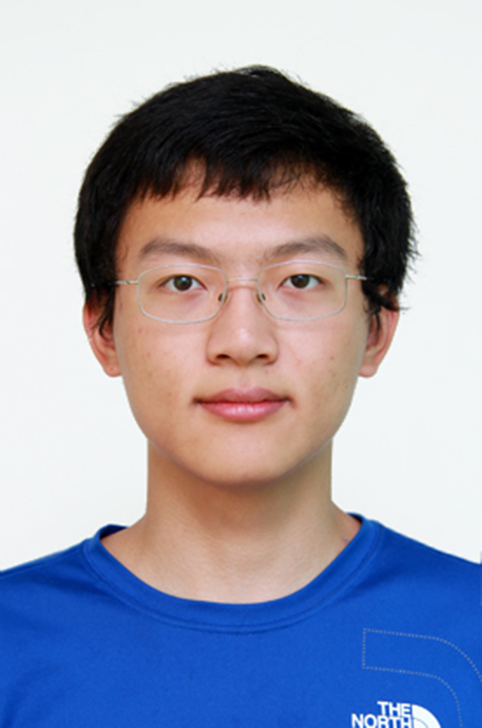}
\textbf{Junzhi Cao} received his PhD degree in Astrophysics and Deep Learning from New York University in 2021. He mainly studies Dialogue System in Natural Language Processing, Large language models, and Cosmology in statistics.
He has published around 10 papers in top journals and refereed conference proceedings, including Nature Communications, Monthly Notices of the Royal Astronomical Society, Journal of Applied Physics, etc.
\endbio

\vskip15pt

\bio{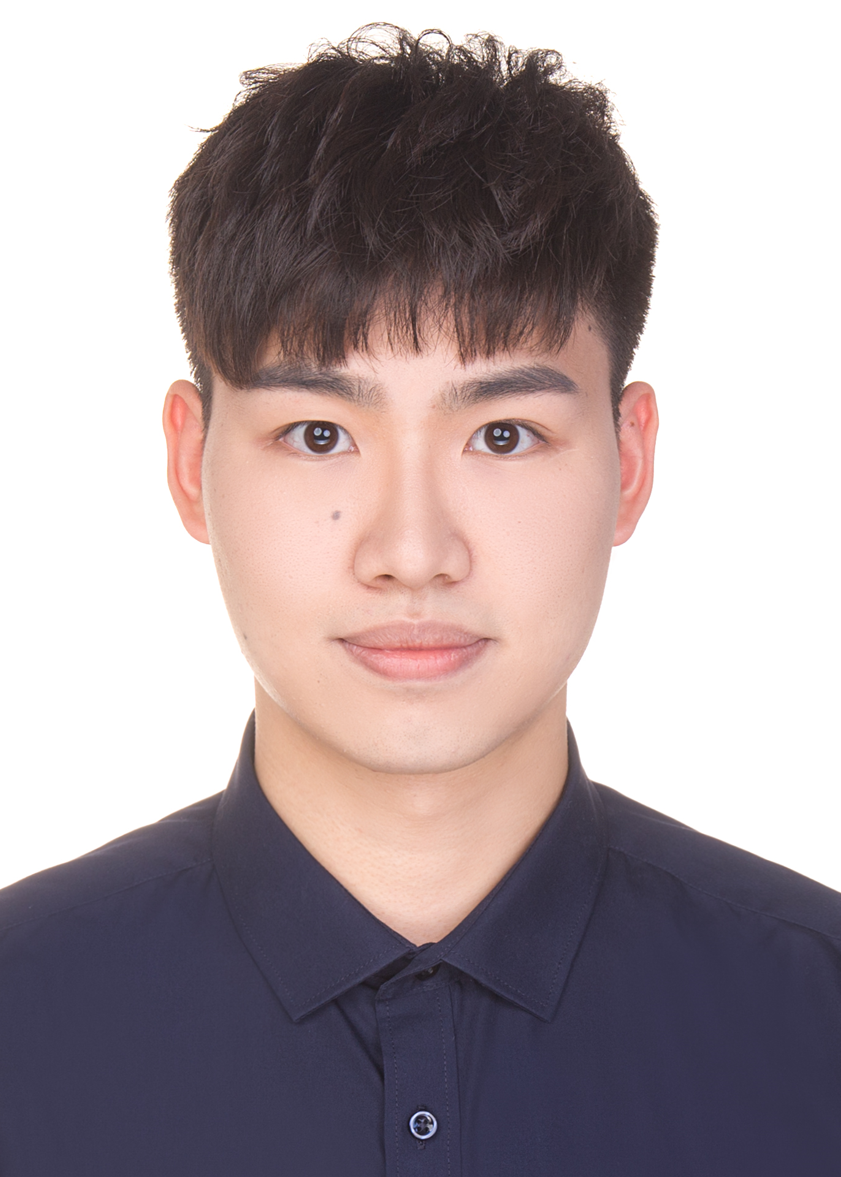}
\textbf{Fengji Zhang} is currently pursuing the master degree with the School of Computer Science, Wuhan University, China. He also received the B.S. degree from the School of Computer Science, Wuhan University in 2020. His current research interests include intelligent software engineering and natural language processing. He has published some papers in top journals and refereed conference proceedings, including IST, JSS, ICLR, and ACL.
\endbio

\vskip70pt

\bio{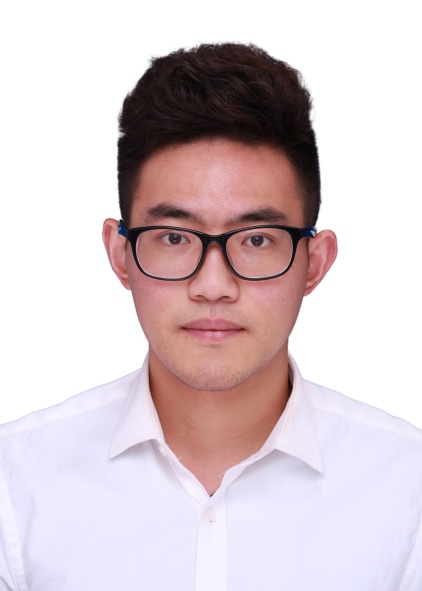}
\textbf{Bingchao Wu} is currently pursuing his Ph.D. degree at the Institute of Software, Chinese Academy of Sciences, Beijing, China. He received his Bachelor's degree in Information Security from Hunan University, Changsha, China in 2017. His research interests include deep learning, recommendation systems, and named entity recognition. He has published several papers in top journals and refereed conference proceedings, including the IEEE ICME, ACL, ISWC, etc.
\endbio


\bio{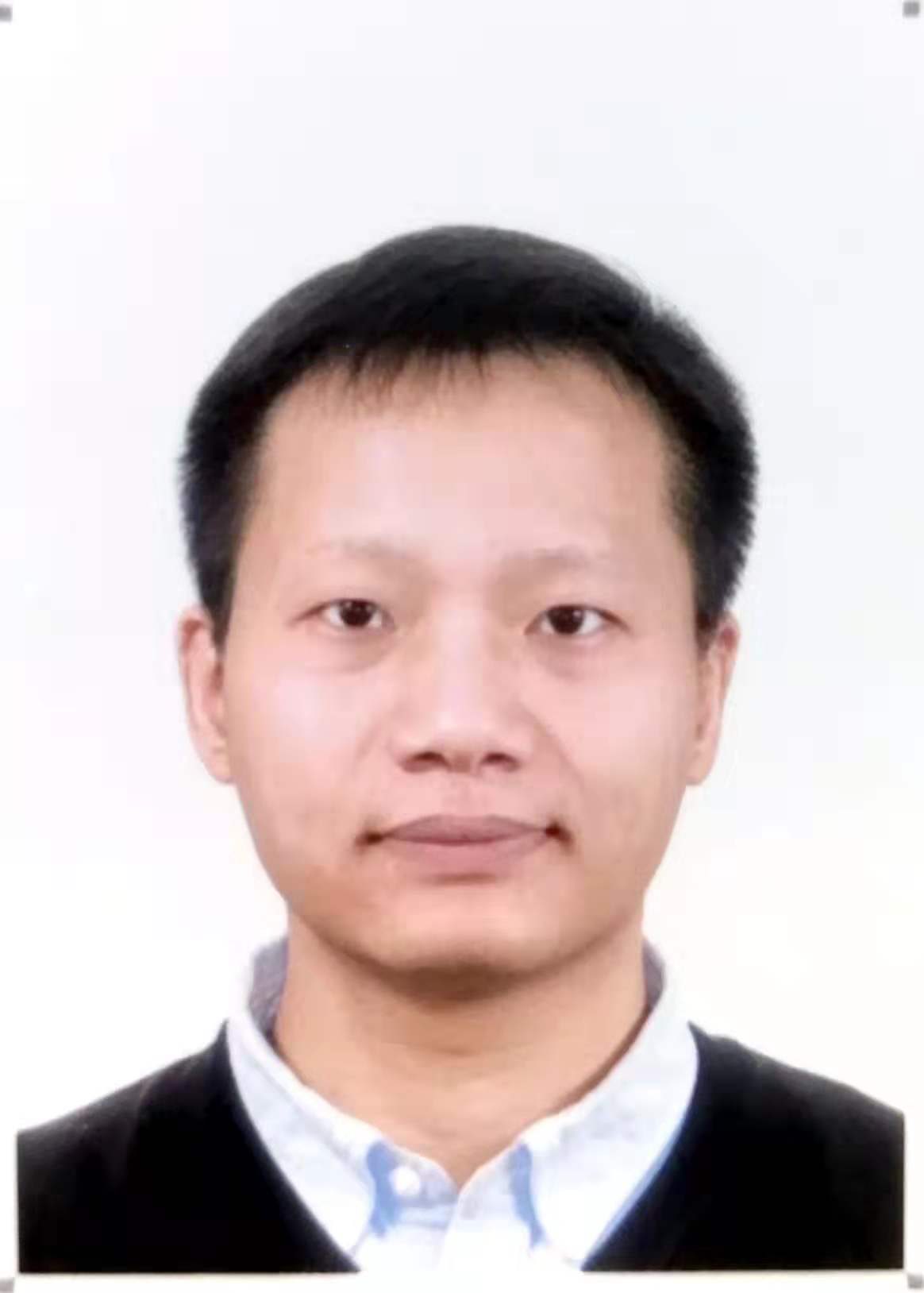}
\textbf{Bei Guan} is currently a Research Associate Professor at Institute of Software, Chinese Academy of Sciences. He received the B.S. degree from Tianjin University in 2007. He got the Ph.D. degree from Institute of Software, Chinese Academy of Sciences in 2015. He got the postdoctoral position at Qatar Computing Research Institute, Hamad Bin Khalifa University (QCRI,HBKU) and finished that in 2018. His primary research interests include big data analytics in healthcare and cyber security, operating system techniques, virtualization techniques, cloud computing, and system security. He has published over 20 technical articles in refereed journals and proceedings, including IEEE Transactions, ACM Transactions, IJCAI, PAKDD, EMNLP, IJCNN, ACL, etc.
\endbio


\bio{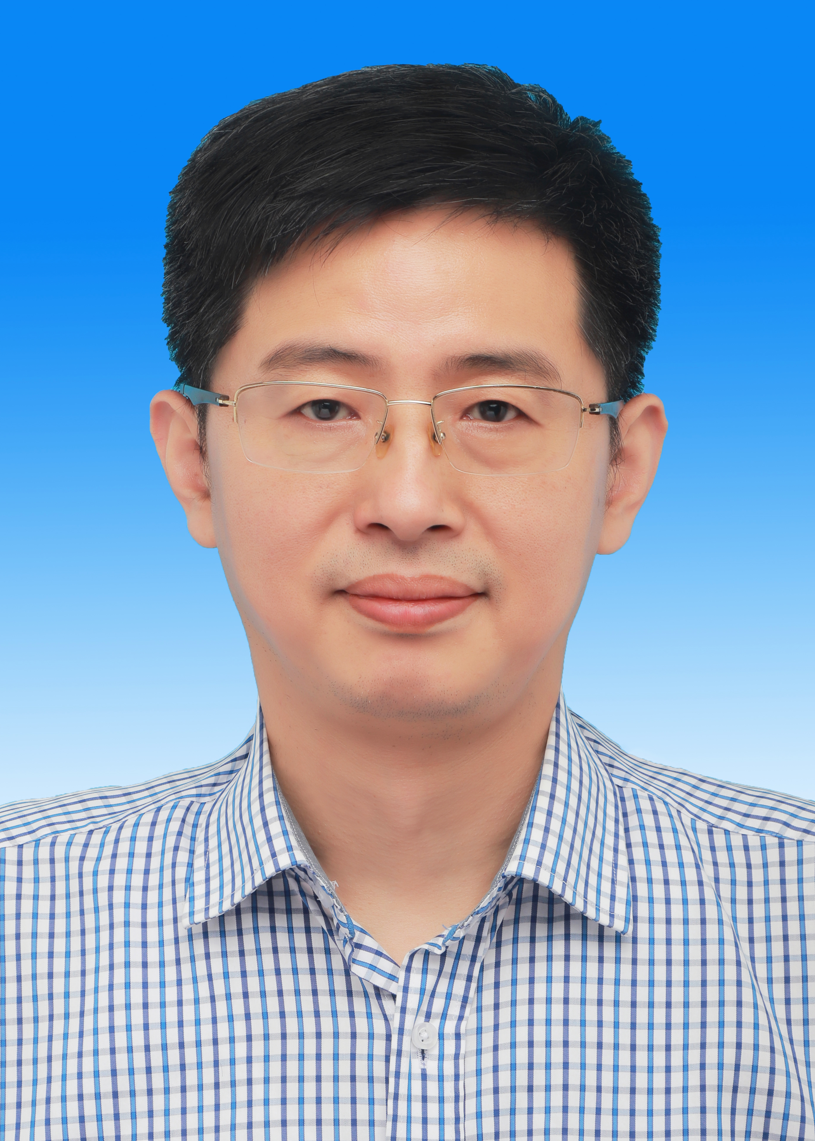}
\textbf{Yilong Yin} is the Director of the Machine Learning and Applications Group and a Distinguished Professor with Shandong University, Jinan, China. He received the Ph.D. degree from Jilin University, Changchun, China, in 2000. From 2000 to 2002, he was a Postdoctoral Fellow with the Department of Electronic Science and Engineering, Nanjing University, Nanjing, China. His research interests include machine learning, data mining, computational medicine, and biometrics. He has published above 100 papers in top journals and refereed conference proceedings, including TKDE, TIP, TMM, ICML, IJCAI, etc.
\endbio

\bio{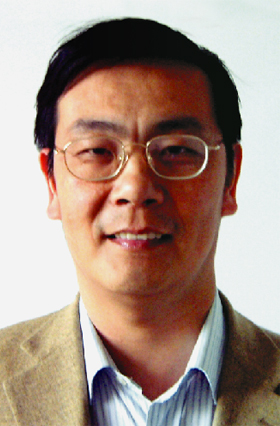}
\textbf{Yongji Wang} is a Distinguished Research Fellow with the Chinese Academy of Sciences and a Ph.D. advisor. He received his Ph.D. degree from the University of Edinburgh, United Kingdom. His research interests include artificial intelligence, big data analysis, and data mining. He has achieved numerous internationally influential research results, participated in over 20 scientific research projects, and published six monographs. He has authored more than 200 high-quality papers in prestigious domestic and international academic journals and conferences, including IEEE Transactions, ACM Transactions, ACL, IJCAI, EMNLP, etc.
\endbio

\end{document}